\documentclass[reprint, prd, nofootinbib, floatfix]{revtex4-2}
\usepackage[utf8]{inputenc}
\usepackage[margin=1in]{geometry}
\usepackage{hyperref}
\usepackage{graphicx}
\usepackage{amssymb}
\usepackage{tikz}
\usepackage{soul}
\usetikzlibrary{graphs, graphs.standard}
\usepackage{stackengine}
\usepackage{multirow}
\usepackage{standalone}
\usepackage{amsmath}
\usepackage{mathtools}
\usepackage{enumitem}
\usepackage{xcolor}
\usepackage{titlesec}

\titlespacing*{\subsection}{0pt}{1.1\baselineskip}{\baselineskip}
\titlespacing*{\subsubsection}{0pt}{1.1\baselineskip}{\baselineskip}

\newcommand{\red}[1]{\textcolor{red}{#1}}

\begin{document}

\title{Creating Simple, Interpretable Anomaly Detectors for New Physics in Jet Substructure}

\author{Layne Bradshaw}
\email{layneb@uoregon.edu}  
\affiliation{Department of Physics and Institute for Fundamental Science, University of Oregon, Eugene, OR 97403}

\author{Spencer Chang}
\email{chang2@uoregon.edu}
\affiliation{Department of Physics and Institute for Fundamental Science, University of Oregon, Eugene, OR 97403}

\author{Bryan Ostdiek}
\affiliation{Department of Physics, Harvard University, Cambridge, MA 02318}
\affiliation{The NSF AI Institute for Artificial Intelligence and Fundamental Interactions}

\date{\today}

\begin{abstract}
    Anomaly detection with convolutional autoencoders is a popular method to search for new physics in a model-agnostic manner. These techniques are powerful, but they are still a ``black box," since we do not know what high-level physical observables determine how anomalous an event is.
    To address this, we adapt a recently proposed technique by Faucett et al., which maps out the physical observables learned by a neural network classifier, to the case of anomaly detection.
    We propose two different strategies that use a small number of high-level observables to mimic the decisions made by the autoencoder on background events, one designed to directly learn the output of the autoencoder, and the other designed to learn the difference between the autoencoder's outputs on a pair of events. 
    Despite the underlying differences in their approach, we find that both strategies have similar ordering performance as the autoencoder and independently use the same six high-level observables.
    From there, we compare the performance of these networks as anomaly detectors.
    We find that both strategies perform similarly to the autoencoder across a variety of signals, giving a nontrivial demonstration that learning to order background events transfers to ordering a variety of signal events.
\end{abstract}

\maketitle
\tableofcontents

\section{Introduction} \label{sec:intro}

Many analyses have been carried out at the Large Hadron Collider (LHC) to look for new physics beyond the Standard Model, but unfortunately these  have yet to yield statistically significant deviations from the expected background. This may indicate that there is no new physics to be found in the data or, more optimistically, it may be a result of not looking for the right signals. There remain many well-motivated models to search for, but designing and carrying out dedicated analyses for each quickly becomes intractable. This motivates the need for broad, model-agnostic searches. The advent of modern machine learning has seen the creation of a variety of unsupervised anomaly detection techniques, all capable of searching for new physics with no reliance on a particular signal model. See Ref.~\cite{Nachman:2020ccu} for a recent review of anomaly detection and unsupervised techniques. 

Anomaly detection techniques rely on an ability to characterize the background in some way, with the hope that this characterization does not generalize to out-of-distribution events, thus making signal events appear ``anomalous." Broadly speaking, anomaly detection can be split into two categories, depending on how similar one expects the signal and background to look. If they are expected to look similar, one has to work to exploit differences in the underlying probability distributions, and many techniques have been developed to highlight those differences~\cite{Kasieczka:2021xcg, Collins:2018epr, DAgnolo:2018cun, DeSimone:2018efk, Casa:2018avf, Dillon:2019cqt, Mullin:2019mmh, DAgnolo:2019vbw, Nachman:2020lpy, Andreassen:2020nkr, ATLAS:2020iwa, Dillon:2020quc, Benkendorfer:2020gek, Mikuni:2020qds, Stein:2020rou, Batson:2021agz, Blance:2020ktp, Bortolato:2021zic, Collins:2021nxn, Dorigo:2021iyy, Volkovich:2021txe, Hallin:2021wme, Buss:2022lxw}. However,  one often expects there to be qualitative differences between signal and background. In that case, there are a variety of methods that can determine whether events are anomalous or not on an event-by-event basis~\cite{Aarrestad:2021oeb, Aguilar-Saavedra:2017rzt, Hajer:2018kqm, Heimel:2018mkt, Farina:2018fyg, Cerri:2018anq, Roy:2019jae, Blance:2019ibf, RomaoCrispim:2019tai, Amram:2020ykb, CrispimRomao:2020ejk, Knapp:2020dde, CrispimRomao:2020ucc, Cheng:2020dal, Khosa:2020qrz, Thaprasop:2020mzp, Aguilar-Saavedra:2020uhm, Pol:2020weg, vanBeekveld:2020txa, Park:2020pak, Chakravarti:2021svb, Faroughy:2020gas, Finke:2021sdf, Atkinson:2021nlt, Dillon:2021nxw, Kahn:2021drv, Caron:2021wmq, Govorkova:2021utb, Gonski:2021jek, Ostdiek:2021bem, Fraser:2021lxm}. 

Machine learning (ML) techniques, including unsupervised anomaly detection, typically make use of low-level, high-dimensional data. This is in contrast to human-engineered strategies, which tend to use high-level, low-dimensional data. When the two perform equally well on a given task, we tend to assume that the ML strategy must have used some combination of its low-level inputs to create an approximation of the high-level variables used by humans. It could be, however, that the ML strategy has found an alternative that is just as efficient. Unfortunately, the ``black box" nature of ML techniques make it difficult to understand what the machine is actually learning. This problem is only amplified when the ML strategy outperforms the human-engineered one. Has the machine learned a simple observable humans didn't consider or has it perhaps found something new?  
 
There have been efforts to understand a neural network by using existing high level observables~\cite{Baldi:2014kfa, DBLP:journals/corr/AlemiFD016, Chang:2017kvc, Wunsch:2018oxb, Roxlo:2018adx}, as well as ``knowledge distillation" techniques to gain insights about complex networks by analyzing simpler ones~\cite{DBLP:journals/corr/abs-2006-05525, Agarwal:2020fpt, Mokhtar:2021bkf, Craven:2020bdz}. In a recent paper (Ref.~\cite{Faucett:2020vbu}), a promising iterative technique was introduced to build an interpretable classifier.  This classifier mimics a ``black box" deep neural network classifier, where the mimicker's inputs consists of a limited set of human-interpretable high-level variables (see also \cite{Collado:2020fwm, Collado:2020ehf}).  In this paper, we extend this technique to anomaly detectors by  presenting two strategies for mapping the low-level information utilized by an anomaly detector into a handful of simple to understand high-level observables.
As a concrete example, we attempt to mimic both the decisions and performance of an anomaly detector based on a convolutional autoencoder, which is trained on background jet images. The convolutional autoencoder then helps to iteratively select high-level observables that serve as the inputs to the mimicker networks. As our pool of high-level observables, we use the Energy Flow Polynomials~\cite{Komiske:2017aww} because they form a basis for all infrared- and collinear-safe observables.

We introduce two strategies to mimic an autoencoder. The first strategy, the \emph{High-Level Network}, uses a small number of high-level observables to match the autoencoder's anomaly score on an event-by-event basis. The other strategy, the \emph{Paired Neural Network}, is tasked with using a potentially different set of observables to learn to make the same ordering decisions as the autoencoder. Given a pair of events, the \emph{Paired Neural Network} learns which of the two was deemed to be less anomalous by the autoencoder. Note that like the convolutional autoencoder we want to mimic, both the \emph{Paired} and \emph{High-Level} neural networks are only trained on background events and so are unsupervised with respect to signal events. Despite their philosophical differences, we find that both strategies agree on which high-level observables are useful for ordering background events like the autoencoder. These two strategies also have comparable performance, where we find that they both make the same ordering decisions as the autoencoder ${\sim} 83\%$ of the time. 

Since these networks are unsupervised, applying these networks as anomaly detectors allows us to test whether the decision ordering on background events transfers to signal events.  Interestingly, for seven of the eight different signals we consider, we find that the mimickers perform as well or better as anomaly detectors than the autoencoder.  Thus, this shows that it is possible to create interpretable anomaly detectors that have a limited number of high-level inputs without compromising performance.  This reduction of complexity is an obvious advantage for experimental applications of anomaly detection, reducing work needed for variable validation and determination of systematic uncertainties.  Theoretically, this result gives insights into the features of a QCD jet image which are harder to compress into a lower dimensional latent space.      

This paper is outlined as follows. In Sec.~\ref{sec:datasets}, we describe the Monte Carlo generated dataset, as well as the relevant selection criteria and preprocessing. Sec.~\ref{sec:methods} starts by describing the details of the convolutional autoencoder. We then review all of the pieces needed to mimic the autoencoder---the pool of high-level observables we use to explain the autoencoder, a metric to determine how similar the decisions of two networks are, the details of our two simplified anomaly detectors, and the iterative procedure we use to construct the mimickers from the pool of high-level observables. We present our results in Sec.~\ref{sec:results}, detailing the construction and performance of the mimickers.  
Finally we conclude in Sec.~\ref{sec:conclusion}.  Details of the simulated events and network training hyperparameters appear in the appendices.

\section{Datasets} \label{sec:datasets}
In this section, we briefly describe the simulated datasets we use in this study. In particular, our focus is on anomaly detection in boosted jets at the LHC.  We utilize the publicly available datasets provided by Ref.~\cite{Cheng:2020dal}, using QCD dijet events \cite{leissner_martin_julien_2020_4641460} as background and $W$, top, and Higgs jets \cite{cheng_taoli_2021_4614656} as the anomalous events. We consider four different $W$ masses, $m_{W}=59, 80, 120, 174$ GeV, two different top masses, $m_{t} = 80, 174\textrm{ GeV,}$ and two different Higgs masses, $m_{h} = 20, 80\textrm{ GeV}$. Note that when $m_{t}=80$ GeV, the mass of the decay product $W$ is set to $20$ GeV. The full simulation details are given in App.~\ref{app:sim_details}.  These signals give a broad range of sig- nals with varying amounts of substructure (two to four prongs), which will prove useful when testing the ability of our anomaly detectors.

These datasets contain approximately $700,000$ QCD dijet events and $100,000$ events for each of the $W$, top, and Higgs signals. After applying a $p_{T}$ cut (see App.~\ref{app:sim_details}), we are left with $\sim 150,000$ QCD events and $\sim 30,0000$ events for each of the anomalous signals We use $2/3$ of the QCD dijet events for training the autoencoder, with the remaining $1/3$ being reserved for testing and validation. We are not considering training on real data at this point, so we do not include the possibility of contamination in the background set from signal samples when training the autoencoder. However, previous work has shown that autoencoders are robust to up to ${\sim}10\%$ signal contamination~\cite{Farina:2018fyg, Heimel:2018mkt, Cerri:2018anq, Mikuni:2021nwn}. 

\begin{figure}
    \centering
    \includegraphics[width=\linewidth]{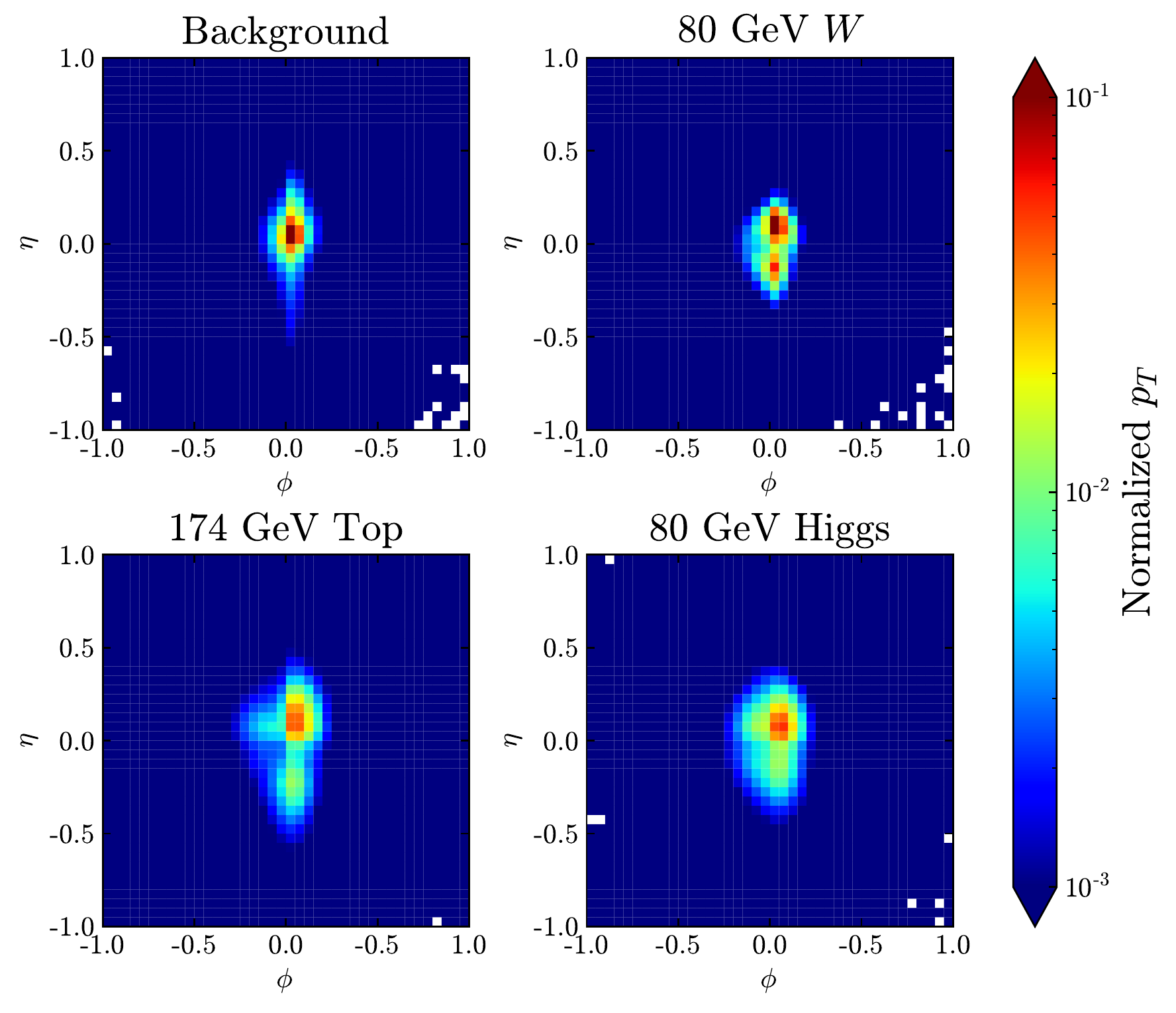}
    \centerline{\begin{minipage}{\linewidth}
    \caption{\small{The average jet image for the background, 80 GeV $W$, 174 GeV top, and 80 GeV Higgs. Note that the Higgs bosons are pair produced from the decay of a heavier Higgs, leading to potentially 4 prongs in the large-radius jet.}}
    \label{fig:avg_img}
    \end{minipage}}
\end{figure}

Our procedure for preprocessing the raw four vectors into images follows that outlined in Ref.~\cite{Macaluso:2018tck} and is implemented with the \textsc{EnergyFlow} package~\cite{energyflow}. For the leading jet in each event, we boost and rotate along the beam direction, such that the $p_{T}$ weighted centroid lies at $\left( \eta, \phi \right) = (0,0)$. The jet is then rotated about its centroid until its principal axis lies along the vertical. Finally, the jet is reflected about the horizontal and vertical axes so that the maximum intensity lies in the upper-right quadrant. Only after centering, rotating, and reflecting the jet do we pixelate the image. Our final pixelated images are $40 \times 40$, covering $\Delta\eta=\Delta\phi=2.0$. The last step of our preprocessing procedure is to divide by the total $p_{T}$ in the image. This final normalization step ensures that each image has the same scale, which helps with training. 
Figure~\ref{fig:avg_img} shows the average jet image for the background and three representative signals---the 80 GeV $W$, 174 GeV top, and 80 GeV Higgs. 

\section{Methodology} \label{sec:methods}

While neural networks have been used for classification and anomaly detection with great success, they are often viewed as black boxes, leading one to wonder what information they are using to match or outperform traditional techniques.
With this in mind, the authors of Ref.~\cite{Faucett:2020vbu} showed that modern classification networks are able to be mimicked by interpretable networks using a few high level physics variables as inputs.
In this work, we adapt this method to the task of anomaly detection.
In order to do this, we first need a good anomaly detector to mimic with physics variables.

\subsection{Creating a Target Anomaly Detector with a Convolutional Autoencoder} 

The anomaly detector we chose is a convolutional autoencoder (hereafter referred to as the AE).
Given an input image, the AE is tasked with encoding the image down into a smaller latent space, then reconstructing the original image from its latent space representation.
The idea behind compressing the data to a smaller representation is that it forces the network to learn what is important about the jet image, while ignoring noisy or less crucial aspects.
The hope is that when the autoencoder is applied to anomalous data, the important characteristics will be different, and thus the image will be poorly encoded, leading to a decoded image which is quite different from the initial image.
Thus, we can distinguish between the background data and the anomalous signal data by the size of the reconstruction error.
AEs were first introduced to the high energy community as anomaly detectors in Refs.~\cite{Hajer:2018kqm,Heimel:2018mkt,Farina:2018fyg}.\footnote{Often, AEs can be improved with Variational Autoencoders (VAEs), in which the latent space representation becomes a distribution, rather than a single point. As a proof of principle, we use the simpler AE, and leave the extension to VAEs for further study.}

\begin{figure}[t]
    \centering
    \includegraphics[width=\linewidth]{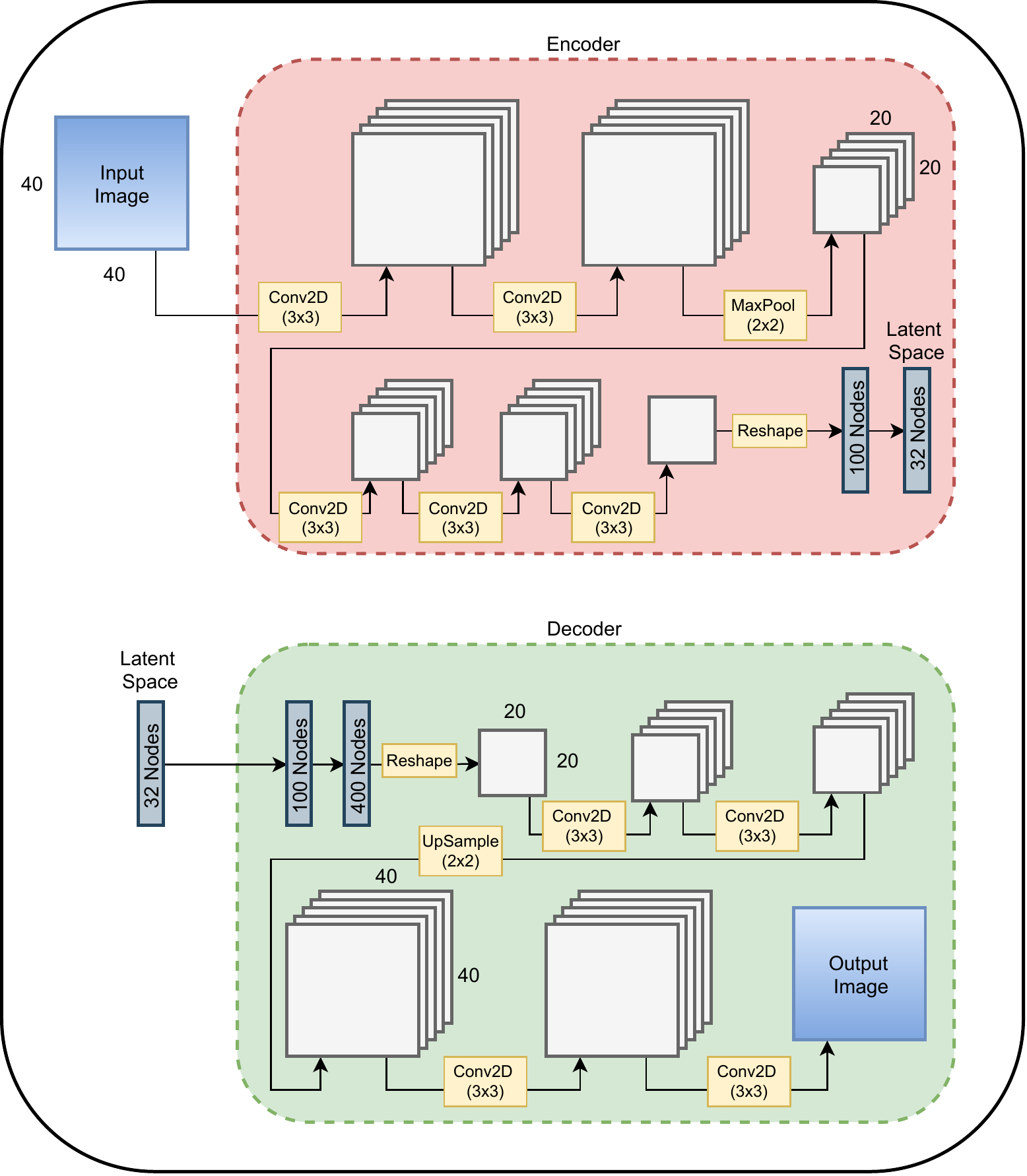}
    \centerline{\begin{minipage}{0.97\linewidth}
    \caption{\small{The architecture of the convolutional autoencoder (AE). The AE consists of two separate networks, an encoder that compresses the original image down to a smaller latent space, and a decoder tasked with recreating the original image from the latent space representation.}}
    \label{fig:ae}
    \end{minipage}}
\end{figure}

The architecture of our AE is shown in Fig.~\ref{fig:ae} and is described below.
The encoder consists of multiple layers.
The first two layers are a set of five $3\times3$ pixel convolutional filters.
We use a stride of one and pad the output to keep the same height and width as the original image.
After each convolution we apply an exponential linear unit (ELU) activation~\cite{clevert2016fast}.
Following these convolutions, the representation is down sampled with a $2\times2$ max pooling layer, leading to a height and width of 20 pixels.
This reduced image is then passed through another two convolutional layers with five filters before being passed through a final convolutional layer with a single filter. This final 20 $\times$ 20 image is then flattened and connected to a Dense layer with 100 nodes, which is in turn connected to our 32-dimensional latent space. We chose a 32-dimensional latent space, as that is where we found the performance of the AE as an anomaly detector began to saturate.

The decoder mirrors the encoder and consists of a Dense layer with 100 nodes, followed by another Dense layer with 400 nodes. Both of these Dense layers use the \textsc{ELU} activation function. The output of this layer is then reshaped into a 20 $\times$ 20 image, and is then passed through two convolutional layers with five filters each. All of the convolutional layers in the decoder use a 3 $\times$ 3 convolutional kernel and the \textsc{ELU} activation function, with the exception of the last convolutional layer in the decoder, which uses the \textsc{Softmax} activation function along the pixel dimension so that the sum of the pixel intensities is unity. These are then upsampled with a transposed convolutional layer to 40 $\times$ 40, passed through a convolutional layer with 5 filters, and finally passed through one last convolutional layer to create the output image.
We train the AE to reproduce QCD jet images,  by minimizing the mean squared error of their reconstruction.
Explicitly, this is given as
\begin{equation}
    L_{\textrm{AE}} = \frac{1}{N_i N_p}\sum_{k}^{N_i} \sum_{\small{j}}^{N_p}\Big(f_{A}(I_{k}^{j}) - I_{k}^{j}\Big)^{2}
    \label{eq:mse_ae}
\end{equation}
where $N_{i}$ is the total number of images, $N_{p}$ is the number of pixels in each image, $I_{k}^{j}$ is the $j$th pixel of the $k$th input image, and $f_{A}(I_{k}^{j})$ is the AE's reconstruction of that pixel for that input image. The training details for the AE are provided in App.~\ref{app:training_details}.
Our AE, along with all of the other neural network architectures discussed in Sec.~\ref{sec:methods} are implemented with \textsc{Keras}~\cite{chollet2015keras} using the \textsc{TensorFlow}~\cite{tensorflow2015-whitepaper} backend.

\begin{figure}[t]
    \centering
    \includegraphics[width=1\linewidth]{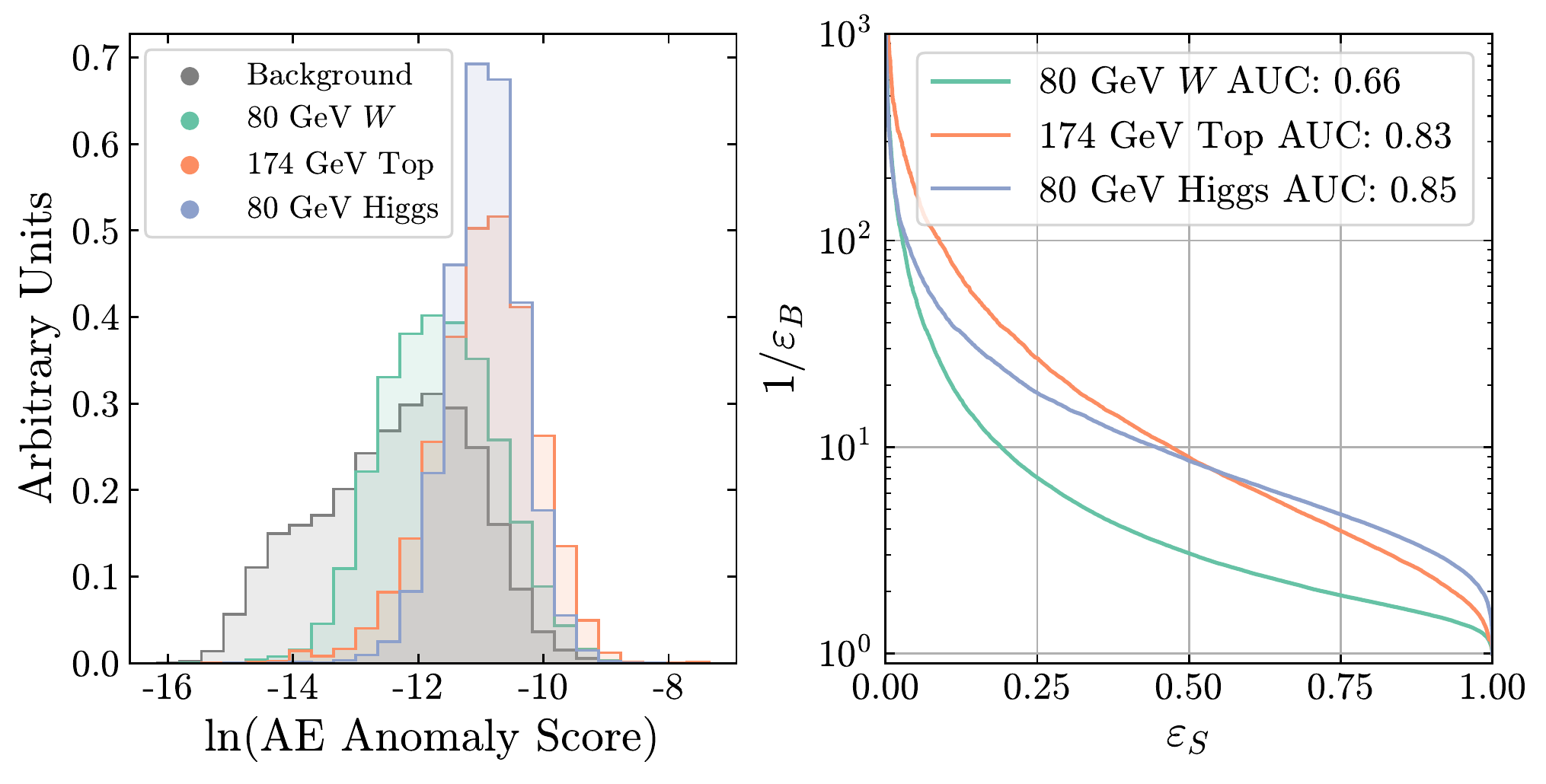}
    \centerline{\begin{minipage}{0.97\linewidth}
    \caption{\small{The AE's performance as an anomaly detector on 3 of the anomalous signals, the 80 GeV $W$, the 174 GeV top, and the 80 GeV Higgs. Note that the Higgs bosons are pair produced from the decay of a heavier Higgs, leading to potentially 4 prongs in the large-radius jet. The left panel shows the normalized distribution of the log of the AE's anomaly score for the background and each of the signals. The right panel shows the ROC curves for each signal.}}
    \label{fig:ae_int}
    \end{minipage}}
\end{figure}
Figure~\ref{fig:ae_int} shows some examples of how the trained AE can act as an anomaly detector.
The left panels display the distribution of the reconstruction errors as the anomaly score for the background training set as well as three different anomalous signals.
At first glance, the reconstruction errors are very small, but this is explained by the normalization and the sparsity of our jet images.
Because each image is normalized to sum to one, all pixels have a value of less than one.
The images are also very sparse, so most pixels are identically 0, and the network is very good at predicting that.
When we take the mean squared error over the pixels, we actually average over the number of pixels, so the number of pixels with no intensity leads to a very good average reconstruction.
Importantly, we see that the background distribution is at lower scores than the signal distributions.
The encoder has never seen jets with inherent substructure from the decay of a heavy resonance, so it doesn't recognize the important information to encode into the latent space, and the decoder therefore performs worse when reconstructing the images.
The right panel displays the Receiver Operating Characteristic (ROC) curves for these three signals.
While the $W$ is harder for the AE to distinguish from the background, the top and Higgs jets have decent Area Under the ROC Curve (AUC) scores.

As we've seen, our constructed AE is capable of detecting jets which are different from the QCD background it was trained on.
In the next section we build up a method to mimic the ordering decisions the AE makes using physics observables.

\subsection{Mimicking the Target Anomaly Detector} 

As shown in the previous section, the AE is able to tag various signals as being different from QCD.
However, it is unclear what information in the event image is being used to do this.
In order to mimic the behavior of the AE, we need a few ingredients.
The first is a wide set of physics observables which could possibly explain the anomaly detector.
For these, we use the Energy Flow Polynomials, described in detail in Sec.~\ref{sssec:efps}.
Next, we use the idea of \emph{decision ordering} to select which observables are important as described in Sec.~\ref{sssec:ado}.
Finally, we need a flexible function which can use the physics observables to produce an anomaly score which mimics that of the AE.
We describe two complementary methods which achieve this goal.
The first method, a Paired Neural Network, is a neural network which takes in the physics observables from two events at the same time and is trained to determine which event had the worse reconstruction error from the AE.
We construct this in such a way that at inference, we can feed in a single event and get an anomaly score.
This technique is described in Sec.~\ref{sssec:pnn}.
The second method, a High-Level Neural Network, instead takes in only a single event at a time and is trained to regress the reconstruction error of the AE for that event.
This second method is described in Sec.~\ref{sssec:hln}.

\subsubsection{High Level Observables} \label{sssec:efps}
 
Since there is no way to know which human-constructed, high-level observables will be relevant \emph{a priori}, we need to rely on using a basis of observables. To that end, we make use of the Energy Flow Polynomials (EFPs)~\cite{Komiske:2017aww}, a formally infinite set of jet substructure observables inspired by previous work on energy correlation functions~\cite{Larkoski:2013eya, Larkoski:2014gra, Banfi:2004yd, Gur-Ari:2011cjr, Jankowiak:2011qa, Moult:2016cvt}.  The EFPs form a discrete linear basis for all infrared- and collinear-safe (IRC-safe) observables and are defined in terms of the momentum fraction, $z_{a}$, and pairwise angular distances, $\theta_{ab}$. The EFPs are computed using the four-momentum of each particle in the jet, where $z_{a}$ is the momentum fraction carried by particle $a$, and $\theta_{ab}$ is the pairwise angular distance between particles $a$ and $b$. Each EFP is conveniently represented by a multigraph, using the following correspondences:
\begin{equation}
    \textrm{each node $a$} \leftrightarrow \displaystyle\sum_{a=1}^{N}z_{a}
\end{equation}
and 
\begin{equation}
    \textrm{each $k$-fold edge between nodes $a$ and $b$} \leftrightarrow \left(\theta_{ab}\right)^{k}.
\end{equation}
As an example, we have
\begin{eqnarray}
        \vcenter{\hbox{\begin{tikzpicture}[transform canvas={scale=0.5}]
		\node[circle, fill=red] (a) at (0:1) {};
		\node[circle, fill=red] (c) at (90:1) {} edge [-] (a);
		\node[circle, fill=red] (d) at (180:1) {} edge [-] (c);
		\node[circle, fill=red] (b) at (270:1) {} edge [-] (c)
									              edge [-, bend left] (a); 
		\draw (d) edge [-, bend left] node {} (c);
		\draw (d) edge [-, bend right] node {} (c);
		\draw (b) edge [-, bend right] node {} (a);
		\node[right] at (a.east) {a};
		\node[left] at (b.west) {b};
		\node[right] at (c.east) {c};
		\node[left] at (d.west) {d};
    \end{tikzpicture}}}
    \qquad = \displaystyle\sum_{a=1}^{N}\sum_{b=1}^{N}\sum_{c=1}^{N}\sum_{d=1}^{N} z_{a}z_{b}z_{c}z_{d}\theta_{ab}^{2}\theta_{ac}\theta_{bc}\theta_{cd}^{3}.
   \nonumber \\[-0.9cm] \label{eq:efp_ex} \\ \nonumber
\end{eqnarray} 
In this example, we've labeled the nodes for clarity, but will not do so for future graphs. To build some intuition for this framework, we note that the fully connected graphs with $N$ vertices correspond to the $N-$point energy correlation functions.

The EFPs corresponding to each multigraph can be modified with a pair of parameters, $\left(\kappa, \beta\right)$, which determine the precise meaning of $z_{a}$ and $\theta_{ab}$. More specifically,
\begin{equation}
    z_{a}^{\left(\kappa\right)} = \left(\frac{p_{T_{a}}}{\sum_{b}p_{T_{b}}}\right)^{\kappa},
\end{equation}
\begin{equation}
    \theta_{ab}^{\left(\beta\right)} = \left(\Delta\eta_{ab}^{2} + \Delta\phi_{ab}^{2}\right)^{\beta/2}
\end{equation}
where $p_{T_{a}}$ is the transverse momentum of particle $a$, $\Delta\eta_{ab}$ is the difference in pseudorapidity between particles $a$ and $b$, and $\Delta\phi_{ab}$ is the difference in azimuthal angle between particles $a$ and $b$. The original IRC-safe EFPs require $\kappa=1$. While there are well-motivated reasons to explore a broader space of observables at the cost of IR and/or C safety~\cite{CMS:2012rth, Larkoski:2014pca, Gras:2017jty}, we restrict ourselves to only IRC-safe observables in this work. For our iterative procedure to mimic the AE, we choose $\kappa=1$, $\beta=1$, and consider all EFPs  with degree (i.e.~the number of edges) $d\leq 5$. With these parameters, we have a total of 102 EFPs to explore.

\subsubsection{Decision Ordering} \label{sssec:ado}
To create an interpretable alternative to the AE, we will iteratively add EFP observables as inputs to the mimicking networks.
To compare how well a network (or EFP input) orders events relative to the AE, we use a series of metrics implemented in Ref.~\cite{Faucett:2020vbu}. Here we briefly summarize these metrics. Given two decision functions, $f(x)\textrm{ and } g(x)$, the \emph{decision ordering} (DO) for a pair of events $x_{1}$ and $x_{2}$ is defined as 
\begin{eqnarray}
    \textrm{DO}\!\left[f, g\right]\!\left(x_{1}, x_{2}\right) = \Theta\Big(\!\big[f(x_{1})-f(x_{2})\big]\!\big[g(x_{1})-g(x_{2})\big]\!\Big)
   \nonumber\\ \label{eq:do} 
\end{eqnarray}
where $\Theta(x)$ is the Heaviside theta function, and we choose $\Theta(0) = 1$.
Here, we can think of $f(x)$ as being the anomaly score/reconstruction error for the AE and $g(x)$ being the output of one of our methods.  Later, we will also use $f(x) = \textrm{AE}(x)$ and $g(x) = \textrm{EFP}(x)$ to determine which EFP observables to include for our mimickers.
A DO of 1 means that $f$ and $g$ agree that one event is more anomalous than another; a DO of 0 indicates the two methods disagree on which event is more anomalous. If two decision functions have $\textrm{DO}=1$ for all possible pairs $x_{1} \textrm{ and } x_{2}$, then the two are effectively identical decision functions on the domain tested.

To create a summary statistic, we then average the DO over all possible pairs, weighted by the underlying distributions that $x_{1}$ and $x_{2}$ are drawn from. The resulting statistic, the \emph{average decision ordering} (ADO) is given by
\begin{equation}
    \textrm{ADO}\!\left[f, g\right] = \int \textrm{d}x_{1}\textrm{d}x_{2}\,p_{1}(x_{1})p_{2}(x_{2})\textrm{DO}\!\left[f, g\right]\!\left(x_{1}, x_{2}\right)
    \label{eq:ado}
\end{equation}
This evaluates to 1 if both decision functions order every possible pair of events in the same manner (making them equivalent decision functions), 0 if they order the pairs in the opposite manner, and $\frac{1}{2}$ if there is no consistency to the way the decision functions order the events. 
Due to computing constraints, we could not compute the ADO on the entirety of the background training set. Instead, when computing the ADO, we choose $10,000$ events at random, and then evaluate on the $\begin{psmallmatrix} 10000 \\ 2\end{psmallmatrix} \sim 5\times 10^{7}$ pairs of events. 

We now follow the \emph{Black-Box Guided Search Strategy} from Ref.~\cite{Faucett:2020vbu} to iteratively construct neural networks whose decision functions should become better and better approximations of the AE's. We start by training a neural network, $\textrm{NN}_{0}$ on some initial set of observables, $X_{0} = (m_J, p_{T})$. We will later describe the two possible architectures for $\textrm{NN}_0$, but for now it is enough to say it aims to produce decision functions that mimic the AE on background events.  We then compute the ADO between $\textrm{NN}_{0}$ and the AE, and isolate all of the pairs of events misordered by $\textrm{NN}_{0}$. From our set of high-level observables, $O$, we then want to find the observable $O_{1}\in O$ with the highest ADO on the pairs misordered by $\textrm{NN}_{0}$.\footnote{If the ADO of an observable is less than 0.5, we take 1-ADO, since a highly anticorrelated variable is also useful.} We then train a new neural network, $\textrm{NN}_{1}$, whose input observables are $X_{1}=X_{0}\cup O_{1}$. Given its inputs, we would expect $\textrm{NN}_{1}$ to have a decision function that more closely resembles that of the AE---and consequently, a higher ADO compared to $\textrm{NN}_{0}$---since it has access to the same information $\textrm{NN}_{0}$ had, as well as information that can help order the pairs misordered by $\textrm{NN}_{0}$. 

From here, we continue to iterate using the remaining observables in  $O$. On the $n$th iteration, we start by finding the observable $O_{n}\in O$ with the highest ADO on the pairs misordered by $\textrm{NN}_{n-1}(X_{n-1})$ that is not already part of $X_{n-1}$. We then build a new set of inputs, $X_{n} = X_{n-1}\cup O_{n}$, and train a new neural network, $\textrm{NN}_{n}$ on $X_{n}$. At each iteration, we expect the ADO between the neural network and AE to increase, since the neural network we construct on the $n$th iteration has access to all of the same information available to the previous network, as well as a new observable $O_{n}$ that helps order the events misordered by the $(n-1)$th neural network.

Now that we have described both the physics observables and the general method for choosing which observables to give the networks, we describe the two network architectures in more detail.

\subsubsection{Paired Neural Network} \label{sssec:pnn}
Our first attempt to mimic the AE is an approach we call the Paired Neural Network (PNN).  The aim of the Paired Neural Network is to mimic the AE by learning to predict the relative anomaly score between two events.
To do this, the PNN takes pairs of events as its input and classifies which has a larger anomaly score.
This is in contrast to other methods such as trying to match the AE's output or anomaly score on an event-by-event basis.
In general, classifiers are easier to train, so this seems like a promising method.

\begin{figure}[t]
    \centering
    \includegraphics[width=0.85\linewidth]{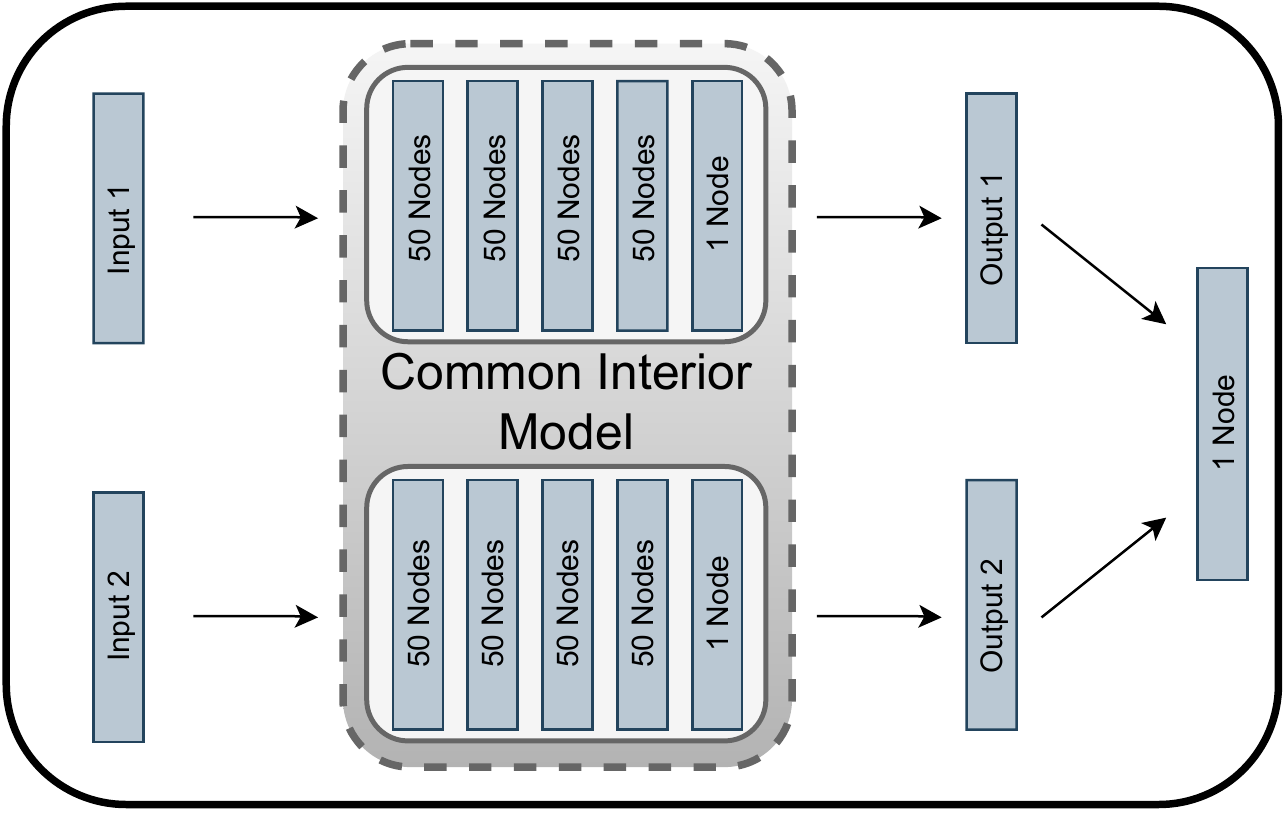}
    \centerline{\begin{minipage}{0.9\linewidth}
    \caption{\small{The architecture of the Paired Neural Network. The interior model consists of 4 hidden layers each with 50 nodes and using the ELU activation function. The interior model outputs a single node for each input and uses the ReLU activation function. The final output of the model is a single node which is the difference between the two interior model outputs and uses a sigmoid activation function. Our input data are the jet's mass, $p_{T}$, and up to 14 EFPs.}} 
    \label{fig:pnn}
    \end{minipage}}
\end{figure}

Figure~\ref{fig:pnn} shows the PNN architecture. 
Both events are fed through the same interior model in parallel. 
This is shown in the image as the ``Common Interior Model."
The interior model consists of four hidden layers with 50 nodes each, and the \textsc{ELU} activation function is used for all layers. 
The interior model produces a single output for each input event, and this single output node uses the \textsc{ReLU} activation. 
The motivation for this is to think of the output for each event as its own anomaly score.
Within the larger PNN, we then subtract these two output anomaly scores from each other.
If the first event is more anomalous, the result should be negative and if the second is more anomalous, the result will be positive.
The larger the difference in scores should tell us about the networks confidence in the relative ordering.
Finally, to turn this into a classification problem, we apply the sigmoid function to the interior model difference, mapping large negative numbers to 0 and large positive values to 1.
If the anomaly scores are the same (the difference is 0) the sigmoid gives a value of 0.5.

To train the network, we continue the idea of classification and minimize the binary cross-entropy given by
\begin{equation}
\begin{split}
    L_{\textrm{PNN}} = -\frac{1}{N}\sum_{k}^{N}\bigg[ & y_{k}\ln\Big(f_{P}(X_{k})\Big) + \\ &(1-y_{k})\ln\Big(1-f_{P}(X_{k})\Big)\bigg]
    \label{eq:bce}
\end{split}
\end{equation}
where $k$ represents a specific pair of events, where the order matters.
The value of $y_{k}$ is the truth ``label" for the pair of events as determined by the AE, i.e. $y_{k}=0\, (1)$ if the AE determines the event in Input 1 to be more (less) anomalous than the event in Input 2, and $f_{P}(X_{k})$ is the PNN's output for the pair of events. Appendix~\ref{app:training_details} provides the training details for the PNN. 
 
After training the PNN on ${\sim}250,000$ pairs of events, we extract the interior model for use on single events.
Thus, even though the training procedure requires pairs of events and was trained as a classifier, the interior model provides a function which takes in observables from a single event and outputs an anomaly score.

\subsubsection{High-Level Neural Network} \label{sssec:hln}
The PNN described in the last section does not attempt to learn the actual anomaly score of the AE, but only the relative difference in the anomaly score between pairs of events.
We also introduce a method which specifically aims to mimic the actual anomaly score of the AE.
We call this network the High- Level Neural Network (HLN).
In practice, the anomaly score (reconstruction error) from the AE spans many orders of magnitude, so we found better results when the HLN is trained to predict the $\log$ of the anomaly score rather than the score itself. 

\begin{figure}[t]
    \centering
    \includegraphics[width=0.8\linewidth]{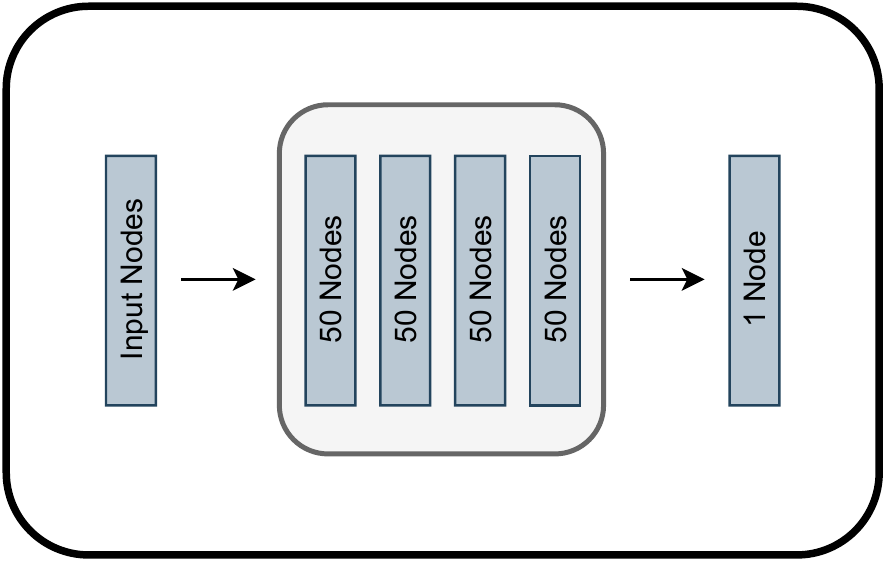}
    \centerline{\begin{minipage}{0.9\linewidth}
    \caption{\small{The architecture of the High- Level Neural Network. This network consists of four hidden layers, with each having 50 nodes and using the ELU activation function. The network output is a single node. Like the PNN, our input data are the jet's mass, $p_{T}$, and up to 14 EFPs.}} 
    \label{fig:hln}
    \end{minipage}}
\end{figure}

We find that a relatively simple neural network is able to achieve the task of reproducing the loss of the AE.  Figure~\ref{fig:hln} shows the architecture we use for the HLN. The HLN consists of 4 hidden layers, with each hidden layer having 50 nodes.
The final output of the network is a single node. 
All of the nodes in the hidden layers use the ELU activation function.

To train the HLN, we minimize the mean squared error between the (log of the) anomaly score of the AE and the output of the HLN.
Specifically, we use a loss function of,
\begin{equation}
 L_{\textrm{HLN}} = \frac{1}{N}\sum_{k}^{N}\Big[f_{H}(X_{k})-\ln \Big(\frac{1}{N_{p}}\sum_{\small{j}}^{N_p}(f_{A}(I_{k}^{j}) - I_{k}^{j})^{2}\Big)\Big]^{2}
 \label{eq:hln_mse}
\end{equation}
where $f_{H}(X_{k})$ is the HLN's output given some input data $X_{k}$ and $f_{A}(I_{k}^{j})$ is the AE's output given a pixel $j$ in an image $I_{k}^{j}$ for the $k$th event. When using the HLN as an anomaly detector, we use the model's output as the anomaly score. See App.~\ref{app:training_details} for the HLN training details.
\section{Results} \label{sec:results}

In the previous section, we outlined two different architectures we could use to iteratively build neural networks whose decision functions would more closely resemble the AE's decision function.  
Here, we provide the results of the iterative procedure and analyze the specific EFPs that are selected to mimic the anomaly detector.
We will find that the EFPs selected are composite observables built out of only six prime EFP factors.
We show that using only the prime components gives very similar results.
Finally, we demonstrate that using the EFPs with a traditional anomaly detection technique, the isolation forest, gives very poor results.
The failure of the isolation forest when provided with the same basic physics observables highlights the benefits of using our mimicker networks.
\subsection{Background Decision Ordering}
\label{sec:results_basemethod}

We start our iterative process by training both a PNN and HLN on jet mass and $p_{T}$ for QCD events in the training set and then compute the ADO for each model.
Of the ${\sim} 5\times 10^{7}$ pairs of events we use to compute the ADO, both the initial PNN and HLN correctly order ${\sim} 72\%$ of the events relative to the reconstruction error of the AE.
Next, we take all of the pairs which are misordered and compute the ADO between all 102 EFPs and the AE.
On this first iteration, we find that the observable with the highest ADO for both networks is EFP 2, given by 
\begin{equation}
  \vcenter{\hbox{\begin{tikzpicture}[transform canvas={scale=0.35}]	\node[circle, fill=red] (a) at (0:1) {};	\node[circle, fill=red] (b) at (180:1) {} edge [-, bend left] (a) edge [-, bend right] (a); \end{tikzpicture}}}\quad\  = \displaystyle\sum_{a,b=1}^{N}z_{a}z_{b}\theta_{ab}^{2}.  
\end{equation}
This observable is then added to the list of inputs. So in the next iteration the input for each event is given by $(m_J, p_T, \rm{EFP~2})$.
We then repeat this process 14 more times, recording both the ADO of each network, as well as which EFP has the largest ADO for the pairs of events which are misordered by the respective networks.

\begin{figure}
    \centering
    \includegraphics[width=0.8\linewidth]{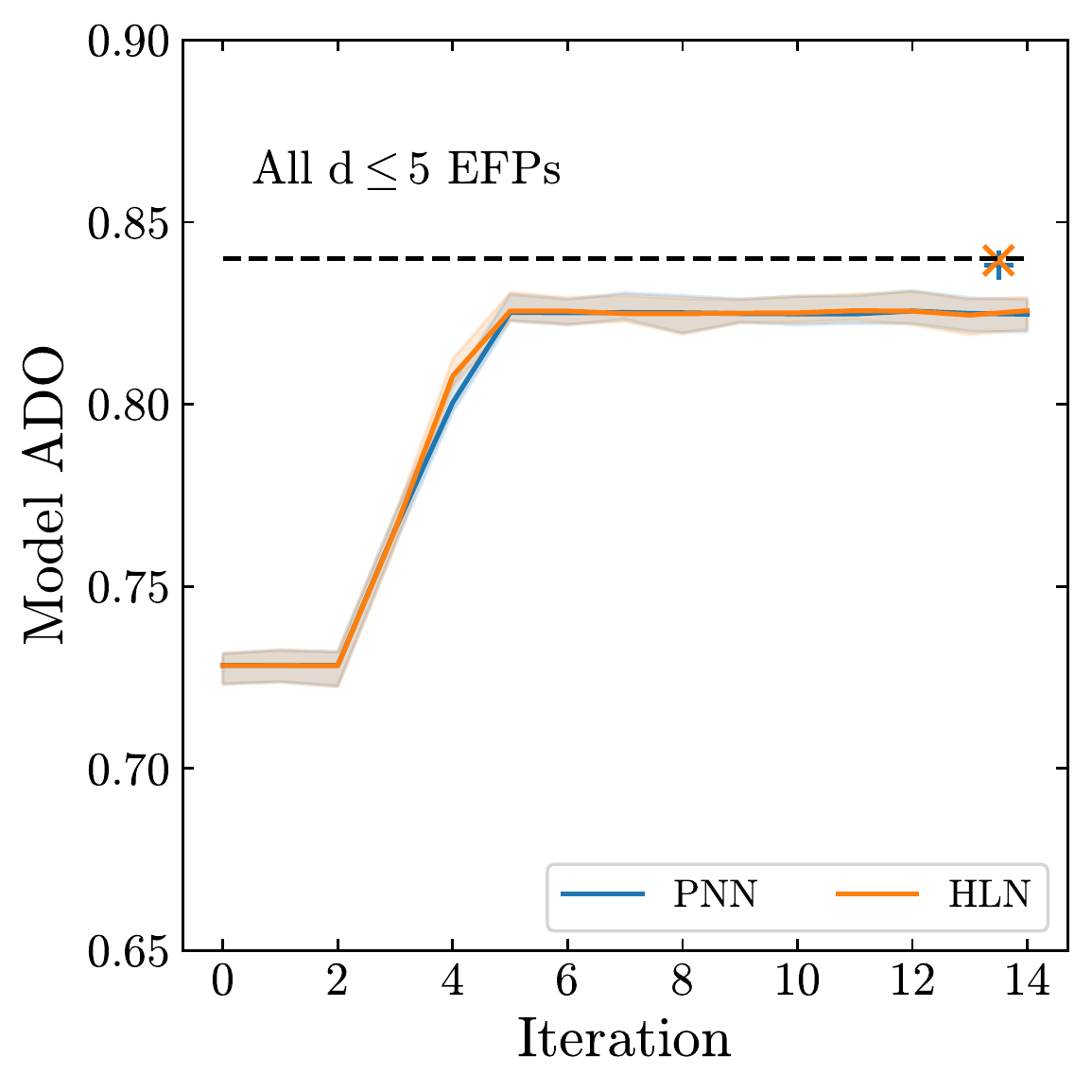}
    \centerline{\begin{minipage}{0.9\linewidth}
    \caption{\small{The ADOs for each PNN and HLN. The center line shows the ADO of the model that was used to select the EFPs. The shaded bands show the maximum and minimum ADO values obtained when recalculating the ADO an additional 50 times, using a different set of pairs of events each time. The $x-$axis denotes the iteration step of the iterative process. See Table~\ref{tab:efps} for the multigraph and mathematical representations of the selected EFPs and the iteration step at which they were added. The blue `$+$' (orange `$\times$') shows the ADO of a PNN (HLN) trained on only the 5 prime EFPs picked out by each method (see Eq.~\ref{eqn:primes}). The ADO of each model trained on $m, p_{T}$, and all of the $d\leq 5$ EFPs is the same to 3 significant digits, and is plotted as a single dashed line.}}  
    \label{fig:ado_2panel}
    \end{minipage}}
\end{figure}

Figure~\ref{fig:ado_2panel} shows the result of this iterative process. 
The solid lines show the ADO of the models we used to determine the next best observable to add; the shaded band shows the maximum and minimum value of the ADO for each model after recalculating the ADO an additional 50 times at each iteration using a different set of ${\sim} 5\times 10^{7}$ pairs of events. 
We also created PNN and HLN models trained on $m, p_{T}$ and all $d\leq 5$ EFPs.  The ADOs of these two models agree to 3 significant digits and thus is plotted as the single dashed line in the panel.  Since they use all of the EFPs, this line gives a sense of the highest ADO each model is capable of achieving, given our set of observables. The blue `$+$' and orange `$\times$' will be discussed in Sec.~\ref{sec:results_prime}.
There are a few key takeaways from these plots.
By the time the ADOs start to plateau, both the HLN and PNN are correctly ordering $83\%$ of the pairs of events in the QCD sample relative to the AE. 
For the first two iterations, the model ADOs do not change. Looking at Table~\ref{tab:efps}, we see that the first two EFPs are EFP 2 and [EFP 2]$^{2}$, which are proportional to $m^{2}/p_{T}^{2}$ and $m^{4}/p_{T}^{4}$. Since the initial inputs to both the PNN and HLN are mass and $p_{T}$, these observables contain no new information, and thus it makes sense that the model ADO does not improve. This redundancy of information follows since the EFPs are a linear basis of substructure observables, whereas our neural networks can utilize nonlinear combinations of its inputs.
Despite their underlying philosophical differences---the HLNs are trying to match the AE's anomaly score, while the PNN is trying the match the DO of the AE---both methods select the same set of 14 EFPs in the same order. In Table~\ref{tab:efps}, we list the multigraph and mathematical expression corresponding to each of these EFPs as well as the iteration step in which they were added.
The agreement of the PNN and HLN approaches gives us confidence that these observables are important to detect jets which do not look like typical QCD jets.  Also, since by the last iteration, the PNN and HLN have nearly reached the ADO of the dashed line, it suggests that the decision ordering of our mimickers has almost converged to what is possible with our set of EFPs.

\begin{table*}[tp]
    \centering
\begin{tabular}{ | c | c | c | c | c |}
    \hline
    EFP No. & EFP Multigraph & EFP Expression & PNN Iteration & HLN Iteration \\
    \hline
    1 & $\vcenter{\hbox{\begin{tikzpicture}[transform canvas={scale=0.35}]
		\node[circle, fill=red] (a) at (0:1) {};
		\node[circle, fill=red] (b) at (180:1) {} edge [-] (a);
    \end{tikzpicture}}}$ & $\displaystyle\sum_{a,b=1}^{N}z_{a}z_{b}\theta_{ab}$ 
    & 5 & 5 \\[12pt]
    2 & $\vcenter{\hbox{\begin{tikzpicture}[transform canvas={scale=0.35}]
		\node[circle, fill=red] (a) at (0:1) {};
		\node[circle, fill=red] (b) at (180:1) {} edge [-, bend left] (a)
									edge [-, bend right] (a);
    \end{tikzpicture}}}$ & $\displaystyle\sum_{a,b=1}^{N}z_{a}z_{b}\theta_{ab}^{2}$ 
    & 1 & 1 \\[12pt]
    54 & $\vcenter{\hbox{\begin{tikzpicture}[transform canvas={scale=0.35}]
        \node[circle, fill=red] (a) at (0:1) {};
        \node[circle, fill=red] (b) at (90:1) {} edge [-] (a);
        \node[circle, fill=red] (c) at (180:1) {};
        \node[circle, fill=red] (d) at (270:1) {} edge [-] (c);
    \end{tikzpicture}}}$ & $\displaystyle\sum_{a,b,c,d=1}^{N}z_{a}z_{b}z_{c}z_{d}\theta_{ab}\theta_{cd}$ 
    & 6 & 6 \\[12pt]
    57 & $\vcenter{\hbox{\begin{tikzpicture}[transform canvas={scale=0.35}]
        \node[circle, fill=red] (a) at (0:1) {};
        \node[circle, fill=red] (b) at (90:1) {} edge [-, bend left] (a)
        							      edge [-, bend right] (a);
        \node[circle, fill=red] (c) at (180:1) {};
        \node[circle, fill=red] (d) at (270:1) {} edge [-, bend left] (c)
        							        edge [-, bend right] (c);
    \end{tikzpicture}}}$ & $\displaystyle\sum_{a,b,c,d=1}^{N}z_{a}z_{b}z_{c}z_{d}\theta_{ab}^{2}\theta_{cd}^{2}$ 
    & 2 & 2 \\[12pt]
    65 & $\vcenter{\hbox{\begin{tikzpicture}[transform canvas={scale=0.35}]
      \node[circle, fill=red] (a) at (0:1) {};
      \node[circle, fill=red] (b) at (72:1) {} edge [-, bend left] (a)
      							    edge[-, bend right] (a);
      \node[circle, fill=red] (c) at (144:1) {};
      \node[circle, fill=red] (d) at (216:1) {} edge [-] (c);
      \node[circle, fill=red] (e) at (288:1) {} edge [-, bend left] (d)
      							      edge [-, bend right] (d);
    \end{tikzpicture}}}$ & $\displaystyle\sum_{a,b,c,d,e=1}^{N}z_{a}z_{b}z_{c}z_{d}z_{e}\theta_{ab}^{2}\theta_{cd}\theta_{de}^{2}$
   & 3 & 3 \\[12pt]
    70 & $\vcenter{\hbox{\begin{tikzpicture}[transform canvas={scale=0.35}]
        \node[circle, fill=red] (a) at (0:1) {};
        \node[circle, fill=red] (b) at (60:1) {} edge [-] (a);
        \node[circle, fill=red] (c) at (120:1) {};
        \node[circle, fill=red] (d) at (180:1) {} edge [-] (c);
        \node[circle, fill=red] (e) at (240:1) {};
        \node[circle, fill=red] (f) at (300:1) {} edge [-] (e);
    \end{tikzpicture}}}$ & $\displaystyle\sum_{a,b,c,d,e,f=1}^{N}z_{a}z_{b}z_{c}z_{d}z_{e}z_{f}\theta_{ab}\theta_{cd}\theta_{ef}$
    & 7 & 7\\[12pt]
    85 & $\vcenter{\hbox{\begin{tikzpicture}[transform canvas={scale=0.35}]
        \node[circle, fill=red] (a) at (0:1) {};
        \node[circle, fill=red] (b) at (60:1) {} edge [-] (a);
        \node[circle, fill=red] (c) at (120:1) {};
        \node[circle, fill=red] (d) at (180:1) {} edge [-, bend left] (c)
        								edge [-, bend right] (c);
        \node[circle, fill=red] (e) at (240:1) {};
        \node[circle, fill=red] (f) at (300:1) {} edge [-, bend left] (e)
        								edge [-, bend right] (e);
    \end{tikzpicture}}}$ & $\displaystyle\sum_{a,b,c,d,e,f=1}^{N}z_{a}z_{b}z_{c}z_{d}z_{e}z_{f}\theta_{ab}\theta_{cd}^{2}\theta_{ef}^{2}$
    & 4 & 4\\[12pt]
    86 & $\vcenter{\hbox{\begin{tikzpicture}[transform canvas={scale=0.35}]
        \node[circle, fill=red] (a) at (0:1) {};
        \node[circle, fill=red] (b) at (51:1) {} edge [-] (a);
        \node[circle, fill=red] (c) at (102:1) {};
        \node[circle, fill=red] (d) at (153:1) {} edge [-] (c);
        \node[circle, fill=red] (e) at (204:1) {};
        \node[circle, fill=red] (f) at (255:1) {} edge [-] (e);
        \node[circle, fill=red] (g) at (306:1) {} edge [-] (a);
    \end{tikzpicture}}}$ & $\displaystyle\sum_{a,b,c,e,d,f,g=1}^{N}z_{a}z_{b}z_{c}z_{d}z_{e}z_{f}z_{g}\theta_{ab}\theta_{ac}\theta_{de}\theta_{fg}$ 
    & 13 & 13 \\[12pt]
    94 & $\vcenter{\hbox{\begin{tikzpicture}[transform canvas={scale=0.35}]
        \node[circle, fill=red] (a) at (0:1) {};
        \node[circle, fill=red] (b) at (51:1) {} edge [-] (a);
        \node[circle, fill=red] (c) at (102:1) {};
        \node[circle, fill=red] (d) at (153:1) {} edge [-] (c);
        \node[circle, fill=red] (e) at (204:1) {};
        \node[circle, fill=red] (f) at (255:1) {} edge [-] (e);
        \node[circle, fill=red] (g) at (306:1) {} edge [-] (a)
                                                  edge [-] (b);
    \end{tikzpicture}}}$ & $\displaystyle\sum_{a,b,c,e,d,f,g=1}^{N}z_{a}z_{b}z_{c}z_{d}z_{e}z_{f}z_{g}\theta_{ab}\theta_{ac}\theta_{bc}\theta_{de}\theta_{fg}$ 
    & 11 & 11 \\[12pt]
    95 & $\vcenter{\hbox{\begin{tikzpicture}[transform canvas={scale=0.35}]
        \node[circle, fill=red] (a) at (0:1) {};
        \node[circle, fill=red] (b) at (45:1) {} edge [-] (a);
        \node[circle, fill=red] (c) at (90:1) {};
        \node[circle, fill=red] (d) at (135:1) {} edge [-] (c);
        \node[circle, fill=red] (e) at (180:1) {};
        \node[circle, fill=red] (f) at (225:1) {} edge [-] (e);
        \node[circle, fill=red] (g) at (270:1) {};
        \node[circle, fill=red] (h) at (315:1) {} edge [-] (g);
    \end{tikzpicture}}}$ & $\displaystyle\sum_{a,b,c,d,e,f,g,h=1}^{N}z_{a}z_{b}z_{c}z_{d}z_{e}z_{f}z_{g}z_{h}\theta_{ab}\theta_{cd}\theta_{ef}\theta_{gh}$ 
    & 8 & 8\\[12pt]
    97 & $\vcenter{\hbox{\begin{tikzpicture}[transform canvas={scale=0.35}]
        \node[circle, fill=red] (a) at (0:1) {};
        \node[circle, fill=red] (b) at (45:1) {} edge [-] (a);
        \node[circle, fill=red] (c) at (90:1) {} edge [-] (b);
        \node[circle, fill=red] (d) at (135:1) {};
        \node[circle, fill=red] (e) at (180:1) {} edge [-] (d);
        \node[circle, fill=red] (f) at (225:1) {};
        \node[circle, fill=red] (g) at (270:1) {} edge [-] (f);
        \node[circle, fill=red] (h) at (315:1) {} edge [-] (a);
    \end{tikzpicture}}}$ & $\displaystyle\sum_{a,b,c,d,e,f,g,h=1}^{N}z_{a}z_{b}z_{c}z_{d}z_{e}z_{f}z_{g}z_{h}\theta_{ab}\theta_{bc}\theta_{cd}\theta_{ef}\theta_{gh}$ 
    & 12 & 12 \\[12pt]
    99 & $\vcenter{\hbox{\begin{tikzpicture}[transform canvas={scale=0.35}]
        \node[circle, fill=red] (a) at (0:1) {};
        \node[circle, fill=red] (b) at (45:1) {} edge [-, bend right] (a)
                                                 edge [, bend left] (a);
        \node[circle, fill=red] (c) at (90:1) {};
        \node[circle, fill=red] (d) at (135:1) {} edge [-] (c);
        \node[circle, fill=red] (e) at (180:1) {};
        \node[circle, fill=red] (f) at (225:1) {} edge [-] (e);
        \node[circle, fill=red] (g) at (270:1) {};
        \node[circle, fill=red] (h) at (315:1) {} edge [-] (g);
    \end{tikzpicture}}}$ & $\displaystyle\sum_{a,b,c,d,e,f,g,h=1}^{N}z_{a}z_{b}z_{c}z_{d}z_{e}z_{f}z_{g}z_{h}\theta_{ab}^{2}\theta_{cd}\theta_{ef}\theta_{gh}$
    & 14 & 14 \\[12pt]
    100 & $\vcenter{\hbox{\begin{tikzpicture}[transform canvas={scale=0.35}]
        \node[circle, fill=red] (a) at (0:1) {};
        \node[circle, fill=red] (b) at (40:1) {} edge [-] (a);
        \node[circle, fill=red] (c) at (80:1) {};
        \node[circle, fill=red] (d) at (120:1) {} edge [-] (c);
        \node[circle, fill=red] (e) at (160:1) {};
        \node[circle, fill=red] (f) at (200:1) {} edge [-] (e);
        \node[circle, fill=red] (g) at (240:1) {};
        \node[circle, fill=red] (h) at (280:1) {} edge [-] (g);
        \node[circle, fill=red] (i) at (320:1) {} edge [-] (a);
    \end{tikzpicture}}}$ & $\displaystyle\sum_{a,b,c,d,e,f,g,h,i=1}^{N}z_{a}z_{b}z_{c}z_{d}z_{e}z_{f}z_{g}z_{h}z_{i}\theta_{ab}\theta_{ac}\theta_{de}\theta_{fg}\theta_{hi}$
    & 10 & 10\\[12pt]
    101 & $\vcenter{\hbox{\begin{tikzpicture}[transform canvas={scale=0.35}]
        \node[circle, fill=red] (a) at (0:1) {};
        \node[circle, fill=red] (b) at (36:1) {} edge [-] (a);
        \node[circle, fill=red] (c) at (72:1) {};
        \node[circle, fill=red] (d) at (108:1) {} edge [-] (c);
        \node[circle, fill=red] (e) at (144:1) {};
        \node[circle, fill=red] (f) at (180:1) {} edge [-] (e);
        \node[circle, fill=red] (g) at (216:1) {};
        \node[circle, fill=red] (h) at (252:1) {} edge [-] (g);
        \node[circle, fill=red] (i) at (288:1) {};
        \node[circle, fill=red] (j) at (324:1) {} edge [-] (i);
    \end{tikzpicture}}}$ & $\displaystyle\sum_{a,b,c,d,e,f,g,h,i,j=1}^{N}z_{a}z_{b}z_{c}z_{d}z_{e}z_{f}z_{g}z_{h}z_{i}z_{j}\theta_{ab}\theta_{cd}\theta_{ef}\theta_{gh}\theta_{ij}$ 
    & 9 & 9\\[12pt]
    \hline
\end{tabular}
\caption{\small{The EFP multigraphs and corresponding expressions for each of the EFPs selected by both the HLN and PNN. In the last two columns, we list the iteration step where the PNN or HLN selects the corresponding EFP.}}
\label{tab:efps}
\end{table*}

\subsection{Anomaly Detection}
\label{sec:results_anomalydetection}

While both the HLN and PNN have demonstrated the ability to mimic the AE's anomaly score on QCD events, it's unclear if matching the decision ordering on in-distribution events will generalize to out-of-distribution events.
In other words, having mimicked the AE on QCD background events with HLNs and PNNs, we must test if this decision ordering transfers to boosted jet signals by comparing the AE, PNN, and HLN as anomaly detectors.
To determine how well each network performs as an anomaly detector we use a popular metric, the AUC.

\begin{figure*}
    \centering
    \includegraphics[width=0.75\linewidth]{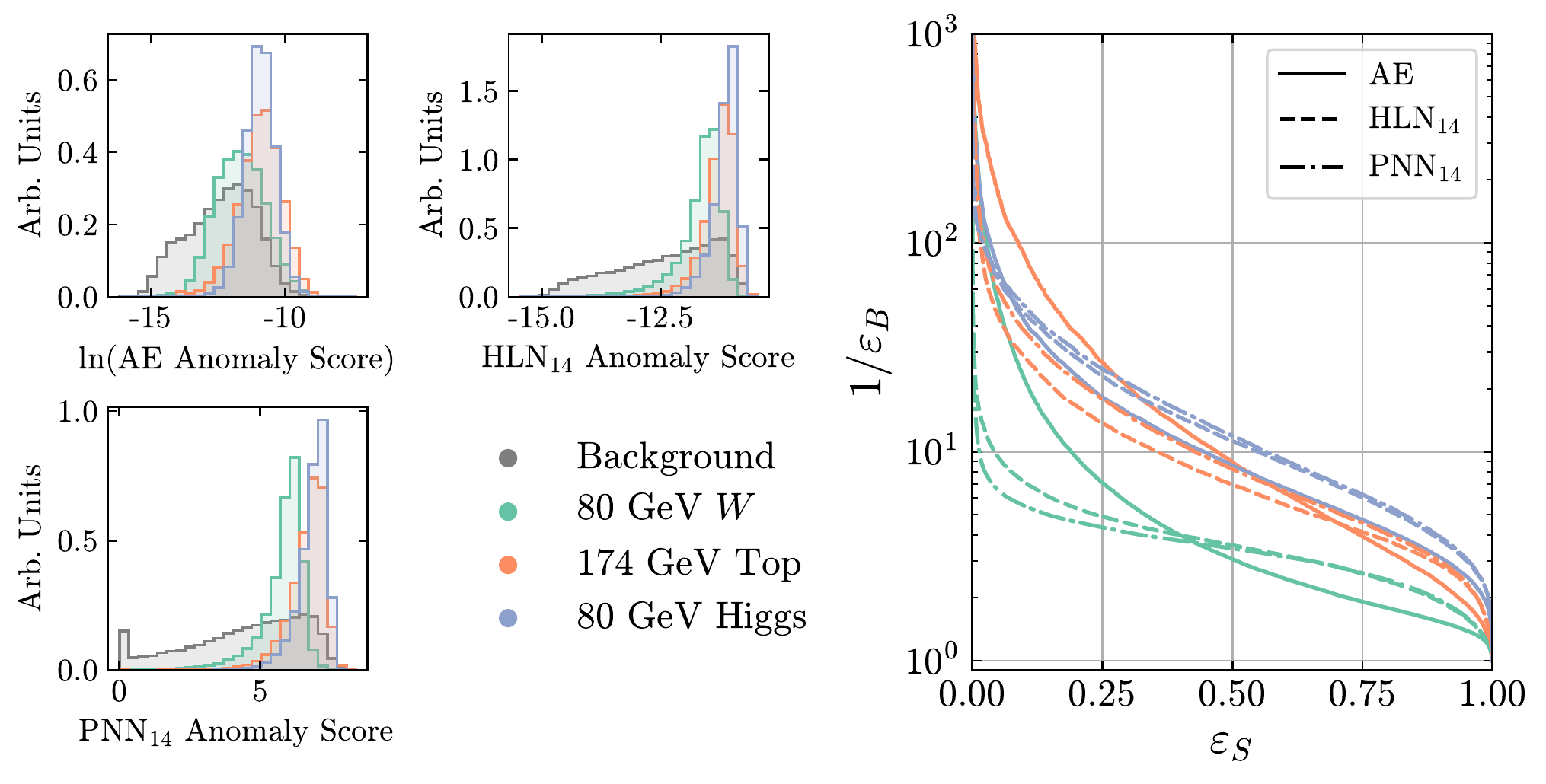}
    \centerline{\begin{minipage}{0.9\linewidth}
    \caption{\small{The performance of the AE, PNN$_{14}$, and HLN$_{14}$ as anomaly detectors on the 80 GeV $W$, 174 GeV top, and 80 GeV Higgs. Note that the Higgs bosons are pair produced from the decay of a heavier Higgs, leading to potentially 4 prongs in the large-radius jet. The left panels show the normalized distribution of each method's respective anomaly score for the background and each signal. The right panel shows the ROC curves for each signal, with the solid lines being the ROC curves for the AE, the dashed lines for HLN$_{14}$, and the dashed-dot lines for PNN$_{14}$.}}
    \label{fig:comp_3panel}
    \end{minipage}}
\end{figure*}

Figure~\ref{fig:comp_3panel} shows how the HLN and PNN on their final iteration compare to the autoencoder on the same three signals as Fig.~\ref{fig:ae_int}. The left panels show the normalized distributions of each network's anomaly scores for the background and three of the signals. The right panel then shows all of the ROC curves for each model on each signal. We can see that both the HLN and PNN do a good job of mimicking the anomaly detector on events with higher anomaly scores. But the long tails in each of the background distributions indicates that HLN and PNN struggle to match the AE on less anomalous events, explaining their poorer background rejection at low signal efficiency. 

\begin{figure*}[t]
    \centering
    \includegraphics[width=\textwidth]{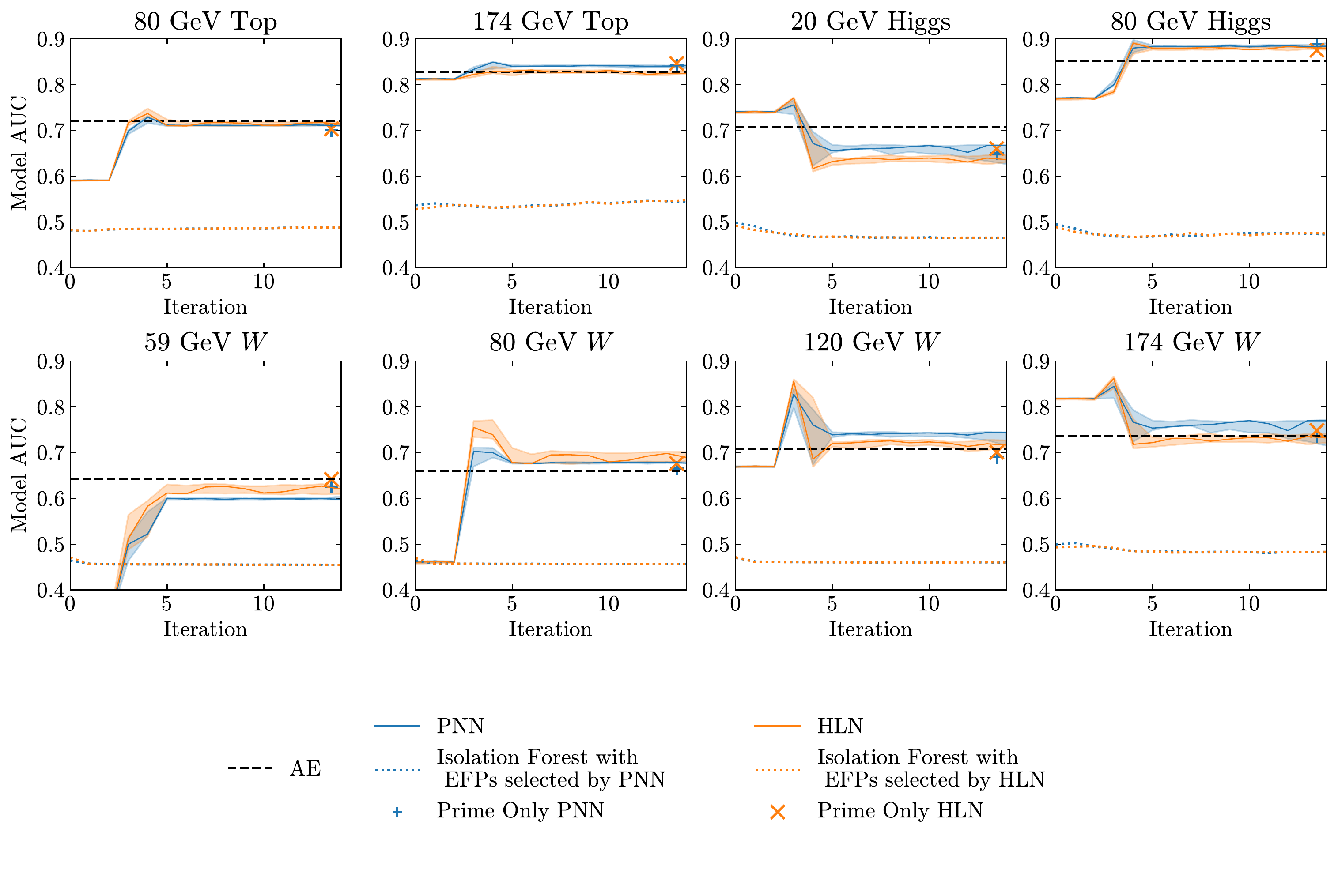}
    \centerline{\begin{minipage}{0.9\linewidth}
    \caption{\small{AUCs for the PNN and HLN at each iteration for each of the eight signals reserved for testing. Note that the Higgs bosons are pair produced from the decay of a heavier Higgs, leading to potentially 4 prongs in the large-radius jet. The solid center lines are the AUC of the model used in the iterative process, the shaded bands show the maximum and minimum AUCs from retraining each network an additional 10 times. The dashed black line corresponds to the AE's AUC. The dotted lines correspond to the isolation forest anomaly detectors and the blue `$+$' (orange `$\times$') is the PNN(HLN) trained using mass, $p_{T}$, and the five prime factors in Eqn.~\ref{eqn:primes}.}}
    \label{fig:auc1}
    \end{minipage}}
\end{figure*}

Figure~\ref{fig:auc1} shows how the mimickers perform on all eight signals described in Sec.~\ref{sec:datasets} at each step of the iterative progress.
The dashed black line in each panel shows the AUC when using the reconstruction error of the AE as the anomaly score.
The blue and orange curves show the results of the PNN and HLN, respectively, as a function of the number of iterations for selecting extra observables.
The solid center lines denote the AUC of the model used to select observables in the iterative process.
The shaded bands show the maximum and minimum AUCs when retraining each network ten additional times, to give us a sense of how stable the training is.
The bands are quite narrow, indicating that the results are robust to training uncertainties.

Like we saw with the ADOs in Fig.~\ref{fig:ado_2panel}, the HLN and PNN perform similarly, despite their different approaches. 
For both the decision ordering and the AUCs, the results start to plateau around the fifth iteration.
When the HLN and PNN AUC scores begin to plateau, we see that the value is similar to the AUC of the AE. This indicates that the HLN and PNN are performing comparably to the AE when all three are acting as anomaly detectors.
It is surprising that mimicking the decision ordering on the in-distribution (QCD) events seems to also generalize to the relative differences between the signals and the background.
Some of the mimicking networks even exceed the anomaly detection capability of the AE they are trying to mimic for certain signals.

For some signals---specifically the 20 GeV Higgs, 80 GeV $W$, 120 GeV $W$, and 174 GeV $W$---we see a drop in AUC around the 3rd iteration for both the PNN and HLN.
While such dips are not ideal, they are not completely unexpected.
Our iterative process is trying to pick out the observables that help to best order the background events, with no attention paid to how effective they may or may not be to picking out signal events.
So, for those three signals, it appears that the EFP added at the iteration where the AUC dips improves the ADO relative to the AE, but at the same time makes it more difficult for the HLNs and PNNs to distinguish those signal events from the  background. 

However, AUC is an inclusive figure of merit and, consequently, does not tell the whole story. As Fig.~\ref{fig:comp_3panel} highlights, networks with similar AUCs are not necessarily making the exact same decisions when used as anomaly detectors. Some more physically interpretable metrics are the background rejection ($1/\varepsilon_{B}$) at fixed signal efficiency ($\varepsilon_{S}$) and the signal efficiency at fixed background rejection. Table~\ref{tab:es_eb-1} shows the background rejection at two different fixed signal efficiencies---0.5 and 0.1---and the signal efficiency at two different fixed values of the background rejection---10 and 100---for all 8 signals and 5 different networks---the AE, HLN$_{0}$, PNN$_{0}$, HLN$_{14}$, and PNN$_{14}$.

\begin{table*}[tp]
\centering
\begin{tabular}{cc}

\begin{tabular}{c}
\textbf{80 GeV Top}
\end{tabular} 
&
\begin{tabular}{c}
\textbf{174 GeV Top}
\end{tabular} \\

\begin{tabular}{ | c | c | c | c | c | c |}
	\hline
	\ & AE & HLN$_{14}$ & PNN$_{14}$ & HLN$_{0}$ & PNN$_{0}$ \\
	\hline
	$\varepsilon_{S}(1/\varepsilon_{B}=10)$ & 0.252 & \red{0.012} & 0.114 & \red{0.071} & \red{0.071}\\
	$\varepsilon_{S}(1/\varepsilon_{B}=100)$ & 0.022 & \red{0.007} & \red{0.008} & \red{0.007} & \red{0.008}  \\
	$1/\varepsilon_{B}(\varepsilon_{S}=0.5)$ & 4.24 & 4.03 & 3.95 & 2.29 & 2.29  \\
	$1/\varepsilon_{B}(\varepsilon_{S}=0.1)$ & 26.5 & 12.0 & 11.3 & \red{7.33} & \red{7.39}  \\
	\hline
\end{tabular}
\hfil & 

\begin{tabular}{ | c | c | c | c | c | c |}
	\hline
	\ & AE & HLN$_{14}$ & PNN$_{14}$ & HLN$_{0}$ & PNN$_{0}$ \\
	\hline
	$\varepsilon_{S}(1/\varepsilon_{B}=10)$ & 0.470 & 0.357 & 0.428 & 0.146 & 0.148 \\
	$\varepsilon_{S}(1/\varepsilon_{B}=100)$ & 0.088 & 0.016 & 0.022 & 0.013 & 0.013  \\
	$1/\varepsilon_{B}(\varepsilon_{S}=0.5)$ & 8.93 & 6.94 & 8.26 & 5.96 & 6.00  \\
	$1/\varepsilon_{B}(\varepsilon_{S}=0.1)$ & 87.6 & 28.6 & 38.0 & 12.8 & 12.9  \\
	\hline
\end{tabular}
\end{tabular}
\medskip

\begin{tabular}{cc}

\begin{tabular}{c}
\textbf{20 GeV Higgs}
\end{tabular} 
&
\begin{tabular}{c}
\textbf{80 GeV Higgs}
\end{tabular} \\

\begin{tabular}{ | c | c | c | c | c | c |}
	\hline
	\ & AE & HLN$_{14}$ & PNN$_{14}$ & HLN$_{0}$ & PNN$_{0}$ \\
	\hline
	$\varepsilon_{S}(1/\varepsilon_{B}=10)$ & 0.240 & \red{0.027} & \red{0.086} & \red{0.032} & \red{0.033} \\
	$\varepsilon_{S}(1/\varepsilon_{B}=100)$ & 0.025 & \red{0.001} & \red{0.001} & \red{0.001} & \red{0.001}  \\
	$1/\varepsilon_{B}(\varepsilon_{S}=0.5)$ & 4.06 & 3.39 & 4.14 & 4.87 & 4.91  \\
	$1/\varepsilon_{B}(\varepsilon_{S}=0.1)$ & 25.7 & \red{6.82} & \red{9.67} & \red{6.68} & \red{6.72}  \\
	\hline
\end{tabular}
\hfil & 

\begin{tabular}{ | c | c | c | c | c | c |}
	\hline
	\ & AE & HLN$_{14}$ & PNN$_{14}$ & HLN$_{0}$ & PNN$_{0}$ \\
	\hline
	$\varepsilon_{S}(1/\varepsilon_{B}=10)$ & 0.446 & 0.549 & 0.565 & \red{0.030} & \red{0.031} \\
	$\varepsilon_{S}(1/\varepsilon_{B}=100)$ & 0.036 & 0.022 & 0.020 & \red{0.002} & \red{0.002}  \\
	$1/\varepsilon_{B}(\varepsilon_{S}=0.5)$ & 8.58 & 11.3 & 11.9 & 4.67 & 4.70  \\
	$1/\varepsilon_{B}(\varepsilon_{S}=0.1)$ & 42.4 & 46.1 & 50.1 & \red{6.41} & \red{6.44}  \\
	\hline
\end{tabular}
\end{tabular}
\medskip

\begin{tabular}{cc}

\begin{tabular}{c}
\textbf{59 GeV $W$}
\end{tabular} 
&
\begin{tabular}{c}
\textbf{80 GeV $W$}
\end{tabular} \\

\begin{tabular}{ | c | c | c | c | c | c |}
	\hline
	\ & AE & HLN$_{14}$ & PNN$_{14}$ & HLN$_{0}$ & PNN$_{0}$ \\
	\hline
	$\varepsilon_{S}(1/\varepsilon_{B}=10)$ & 0.155 & \red{0.017} & \red{0.007} & \red{0.011} & \red{0.012} \\
	$\varepsilon_{S}(1/\varepsilon_{B}=100)$ & 0.015 & \red{0.0003} & \red{0.0003} & \red{0.0007} & \red{0.0007}  \\
	$1/\varepsilon_{B}(\varepsilon_{S}=0.5)$ & 2.86 & 2.76 & 2.62 & \red{1.40} & \red{1.40}  \\
	$1/\varepsilon_{B}(\varepsilon_{S}=0.1)$ & 16.1 & \red{5.08} & \red{3.91} & \red{2.36} & \red{2.35}  \\
	\hline
\end{tabular}
\hfil & 

\begin{tabular}{ | c | c | c | c | c | c |}
	\hline
	\ & AE & HLN$_{14}$ & PNN$_{14}$ & HLN$_{0}$ & PNN$_{0}$ \\
	\hline
	$\varepsilon_{S}(1/\varepsilon_{B}=10)$ & 0.190 & \red{0.043} & \red{0.013} & \red{0.014} & \red{0.014} \\
	$\varepsilon_{S}(1/\varepsilon_{B}=100)$ & 0.028 & \red{0.0005} & \red{0.0004} & \red{0.0009} & \red{0.0009}  \\
	$1/\varepsilon_{B}(\varepsilon_{S}=0.5)$ & 3.06 & 3.57 & 3.44 & \red{1.77} & \red{1.77}  \\
	$1/\varepsilon_{B}(\varepsilon_{S}=0.1)$ & 22.4 & \red{7.17} & \red{5.52} & \red{2.83} & \red{2.84}  \\
	\hline
\end{tabular}
\end{tabular}
\medskip

\begin{tabular}{cc}

\begin{tabular}{c}
\textbf{120 GeV $W$}
\end{tabular} 
&
\begin{tabular}{c}
\textbf{174 GeV $W$}
\end{tabular} \\

\begin{tabular}{ | c | c | c | c | c | c |}
	\hline
	\ & AE & HLN$_{14}$ & PNN$_{14}$ & HLN$_{0}$ & PNN$_{0}$ \\
	\hline
	$\varepsilon_{S}(1/\varepsilon_{B}=10)$ & 0.244 & \red{0.070} & \red{0.089} & \red{0.021} & \red{0.022} \\
	$\varepsilon_{S}(1/\varepsilon_{B}=100)$ & 0.040 & \red{0.001} & \red{0.001} & \red{0.001} & \red{0.001}  \\
	$1/\varepsilon_{B}(\varepsilon_{S}=0.5)$ & 3.71 & 4.01 & 4.76 & 2.97 & 2.97  \\
	$1/\varepsilon_{B}(\varepsilon_{S}=0.1)$ & 32.9 & \red{8.52} & \red{9.58} & \red{4.30} & \red{4.31}  \\
	\hline
\end{tabular}
\hfil & 

\begin{tabular}{ | c | c | c | c | c | c |}
	\hline
	\ & AE & HLN$_{14}$ & PNN$_{14}$ & HLN$_{0}$ & PNN$_{0}$ \\
	\hline
	$\varepsilon_{S}(1/\varepsilon_{B}=10)$ & 0.289 & 0.124 & 0.190 & \red{0.064} & \red{0.064} \\
	$\varepsilon_{S}(1/\varepsilon_{B}=100)$ & 0.052 & \red{0.003} & \red{0.003} & \red{0.003} & \red{0.003}  \\
	$1/\varepsilon_{B}(\varepsilon_{S}=0.5)$ & 4.40 & 4.21 & 5.53 & 6.05 & 6.10  \\
	$1/\varepsilon_{B}(\varepsilon_{S}=0.1)$ & 42.4 & 11.4 & 14.7 & \red{8.61} & \red{8.57}  \\
	\hline
\end{tabular}
\end{tabular}
\medskip

\caption{\small{The background rejection ($1/\varepsilon_{B}$) at two different fixed signal efficiencies ($\varepsilon_{S}$)---0.5 and 0.1---and the signal efficiency at two different fixed values of the background rejection---10 and 100---for all 8 anomalous signals. We present these metrics for 5 different networks, the AE, PNN$_{0}$, HLN$_{0}$, PNN$_{14}$, and HLN$_{14}$. The values shown in red are those where $\varepsilon_{B} > \varepsilon_{S}.$}}
\label{tab:es_eb-1}
\end{table*}

There are a few key takeaways from this table. Looking at the signal efficiency at a fixed value of the background rejection, we can see that, in general, our mimicker networks need to operate at lower signal efficiencies to achieve the same background rejection as the AE. The exceptions here are the final iteration of the mimicker networks when used as anomaly detectors for the 174 GeV Top and 80 GeV Higgs. These networks, when applied to these signals operate at comparable signal efficiencies to the AE for lower fixed values of the background rejection. Shifting now to the background rejection at fixed signal efficiency, we see that our mimicker networks compare favorably to the AE at higher signal efficiencies across all of the anomalous signals we consider, but fall behind the AE at lower signal efficiencies. Again, the exception here are the mimicker networks applied to the 80 GeV Higgs. As was observed earlier in Fig.~\ref{fig:comp_3panel}, as we make tighter cuts on our mimicker networks, forcing them to operate at lower signal efficiencies, they begin to deem the background as being more anomalous than the signal when compared to the autoencoder. While this type of behavior would be difficult to deal with in a real analysis, it is not unique to our mimicker networks and is a challenge with anomaly detection in general. The cuts that result in $\varepsilon_{B} > \varepsilon_{S}$ are highlighted in red in Tab.~\ref{tab:es_eb-1}. Taken together, these indicate that most of the performance of our mimicker networks is coming at higher signal efficiencies, and the long tails in their anomaly scores for the background distribution holds them back from exactly matching the AE.

Finally, by the endpoint of the iterative process, we had found that the PNN and HLN agreed on ordering of background events at about $83\%$ when compared to the AE.  Here, we see that in terms of the AUC metric, $83\%$ mimicking transferred quite well to the use of these mimickers as simpler anomaly detectors with comparable performance. We expect the tendency for the mimicker networks to tag the background as being more anomalous than the signal at low signal efficiencies to subside as the ADO of the mimickers approaches 1.

\subsection{Using Only Prime EFPs}
\label{sec:results_prime}

In examining the EFPs selected to improve the decision ordering, we note that even though we use up to 14 EFPs, they only depend on six prime EFP factors:
\begin{equation}
\vcenter{\hbox{\begin{tikzpicture}[transform canvas={scale=0.35}]	\node[circle, fill=red] (a) at (90:1) {};	\node[circle, fill=red] (b) at (270:1) {} edge [-] (a); \end{tikzpicture}}}\ ,
\quad\quad\vcenter{\hbox{\begin{tikzpicture}[transform canvas={scale=0.35}]	\node[circle, fill=red] (a) at (90:1) {};	\node[circle, fill=red] (b) at (270:1) {} edge [-, bend right] (a) edge [-, bend left] (a); \end{tikzpicture}}}\ \, , 
\quad \quad \vcenter{\hbox{\begin{tikzpicture}[transform canvas={scale=0.35}]	\node[circle, fill=red] (a) at (0:1) {};	\node[circle, fill=red] (b) at (90:0) {} edge [-] (a); \node[circle, fill=red] (c) at (180:1) {} edge [-] (b); \end{tikzpicture}}}\quad\, ,
\quad \quad \vcenter{\hbox{\begin{tikzpicture}[transform canvas={scale=0.35}]	\node[circle, fill=red] (a) at (90:1) {};	\node[circle, fill=red] (b) at (210:1) {} edge [-] (a); \node[circle, fill=red] (c) at (330:1) {} edge [-, bend left] (b) edge[-, bend right] (b); \end{tikzpicture}}}\quad\, ,
\quad \quad \vcenter{\hbox{\begin{tikzpicture}[transform canvas={scale=0.35}]	\node[circle, fill=red] (a) at (90:1) {};	\node[circle, fill=red] (b) at (210:1) {} edge [-] (a); \node[circle, fill=red] (c) at (330:1) {} edge [-] (b) edge [-] (a); \end{tikzpicture}}}\quad\, ,
\quad \quad \vcenter{\hbox{\begin{tikzpicture}[transform canvas={scale=0.35}] \node[circle, fill=red] (a) at (180:1) {}; \node[circle, fill=red] (b) at (180:0.33) {} edge (a); \node[circle, fill=red] (c) at (0:0.33) {} edge [-] (b); \node[circle, fill=red] (d) at (0:1) {} edge [-] (c); \end{tikzpicture}}} 
\label{eqn:primes}
\end{equation}
Notably in these primes, the first and fifth prime factors are the energy correlation functions for two and three prong structures \cite{Larkoski:2013eya}. It is also  interesting to note that these prime factors are nonzero only for $\geq 2, 3$ prong structures.
As the AE is learning to encode the predominantly 1-prong QCD events, it seems that it is losing information contained in these higher prong observables.
With this loss of information, networks with direct access to these observables are able to explain the reconstruction error of the network.

The observation that the anomaly scores can be explained by composite operators which only have a few prime operators leads one to wonder if the prime EFPs are good enough.
To test this, we trained both the PNN and the HLN using mass, $p_T$, and the six prime EFPs.
The results are denoted in Fig.~\ref{fig:ado_2panel} and Fig.~\ref{fig:auc1} by the blue `$+$' and orange `$\times$', respectively.
Not only do these ``prime-only" networks perform comparably to each other, which matches the behavior we saw from the networks trained on the composite EFPs, but the prime-only and composite networks also perform comparably across all of the signals.
The results in Fig.~\ref{fig:ado_2panel} show the ADO of the prime-only networks computed on the same pairs of events as the center line for the composite models. The ADO of the prime-only models has a similar spread as the composite models, and thus the two do indeed perform comparably. 
Taken together, this seems to indicate that the prime EFPs alone contain all of the necessary information to construct simple anomaly detectors capable of matching much more complex ones.
While each of the prime EFPs on their own would have been selected eventually, these results also suggests a more efficient iterative procedure for creating HLN and PNN mimickers, where one uses the redundancy in the full space of EFPs to their advantage and allows the algorithm to explore the full space of composite EFPs, but only selects those containing new prime factors. 
\subsection{Comparison with Isolation Forests}
\label{sec:results_iso}
Through this iterative process, we've constructed two different types of dense neural networks that approximately match the AE not only in how their decision functions order background events, but also as anomaly detectors for classifying a variety of signals. It is clear then that the observables picked out by this procedure contain the information needed to match the AE on both fronts. One then wonders if an even simpler anomaly detector than the ones presented in  Sec.~\ref{sec:methods} would give similar results. To investigate this possibility, we consider isolation forests as implemented by \textsc{Isolation Forest} in \textsc{SciKit-Learn}~\cite{scikit-learn}.

Isolation forests work by randomly selecting a feature from a given set of inputs, and then randomly selecting a split value for that feature. This splitting process is repeated until each event the model is trained on has been isolated from the rest, resulting in a tree-like structure. We then build an ensemble, or ``forest" of these classifiers. The anomaly score is the number of splittings needed to isolate each event, averaged over the entire ensemble. This kind of random partitioning tends to take fewer splittings to isolate anomalous events, so if the average number of splittings across a large ensemble is low, the event is likely to be anomalous. We wanted to see if the performance of the isolation forests saturate in the same way the HLNs and PNNs did, so we trained a series of them and added the new observable picked out by either the HLN or PNN each time. The details of our specific implementation is given in App.~\ref{app:training_details}. Since the HLNs and PNNs selected EFPs in a slightly different order, we trained 2 different sets of isolation forests. One set added observables in the order selected by the HLN, while the other added them in the order selected by the PNN. 

Figure~\ref{fig:auc1} shows how the isolation forests compare to the HLNs, CNNs, and AE when used as a classifier on the 8 signals considered in this work.
The blue dotted line shows the AUC of the isolation forests trained on the EFPs selected by the PNN, the orange dotted line corresponds to isolation forests trained on the EFPs selected by the HLN.
For most of the signals, both isolation forests have an AUC of ${\sim} 0.5$, and are unable to match the performance of the HLN, PNN, or AE.
This is a very interesting observation.
The same small set of observables are able to lead to good anomaly detection when trying to match the decisions of the AE.
However, as discussed above, these observables in some sense tell us what the AE is choosing to ignore when learning to reconstruct QCD images.
Since these observables are not very descriptive for QCD events, the isolation forest does not have much to learn from.
We expect the results would hold for other anomaly detection techniques trained on the same observables.
Thus, we suspect it is the mimicking aspect of our procedure which allows for good anomaly detection with the simple set of observables.
\section{Conclusion} \label{sec:conclusion}
In this paper, we have extended the results of Ref.~\cite{Faucett:2020vbu} to build simpler, more interpretable anomaly detectors.  Starting with a convolutional autoencoder, we iteratively built a network that mimics the autoencoder's ordering of background events, where the network's inputs are high-level variables taken from a set of Energy Flow Polynomials.  We presented two network architectures for the mimickers, the High-Level Network and the Paired Neural Network.  The High-Level Network aims to reproduce the reconstruction error of the autoencoder, while the Paired Neural Network takes in two events and is trained to order them like the autoencoder. Note that both the PNN and HLN are trained to order anomalous events from the physics observables, which is an inherently different task than the autoencoder, which was only trained to compress and decompress background data. This highlights the difference with Ref.~\cite{Faucett:2020vbu}, in which the black-box network and mimicking network have the same task of binary classification. Given this fundamental difference between our AE and mimicking networks, it is not obvious that employing the same strategy will work when trying to mimic the autoencoder's ordering. However, we find that these two complementary approaches give similar performance, ${\sim} 83\%$ agreement, when ordering background events and also pick out the same list of EFPs, suggesting the commonality of the information that is needed to order events like the autoencoder.

After mimicking the autoencoder on ordering of background events, we take these networks and apply them as anomaly detectors on eight different signals.  Even though the mimickers and autoencoder have never seen these events, we find that the similarity in ordering transfers to these events, making the mimickers as good (or better) than the autoencoder as an anomaly detector for seven of the eight signals.  It is worth emphasizing how such results were not guaranteed to occur. The autoencoder, having been trained only on background events, has no concept of what is anomalous.
So it is not obvious that mimicking the ordering of events for the background will generalize to anomalous events, especially given a large set of signal classes. 

Since the high-level observables picked out by these mimickers rely only on six prime Energy Flow Polynomials, it indicates that the information required to order events like the autoencoder is reasonably small.  However, since the isolation forests based on these high-level inputs did not perform as well, it shows that mimicking the autoencoder's background ordering is crucial in creating a simpler anomaly detector.

In terms of future directions, it would be interesting to extend the list of Energy Flow Polynomials to check that one can saturate the decision ordering of the autoencoder and to determine what prime Energy Flow Polynomials are needed for that.  Applying this technique to other anomaly detection methods on the same dataset would help uncover what high-level variables are being used by these methods and could help in designing more powerful anomaly detectors.  Finally, it would be interesting to see if one can extend this technique to cases where there is no known high-level variable basis (like the Energy Flow Polynomials) and to see to what extent decision ordering transfers to different signals.
For instance, the methods which performed best on the Dark Machines anomaly score challenge~\cite{Aarrestad:2021oeb,Caron:2021wmq,Ostdiek:2021bem} used variational autoencoder structures which only aimed to make a Gaussian latent space and did not try to reconstruct events.
It would be very interesting to see what physics these methods are using, but there is no obvious basis of observables to use.

\section*{Acknowledgements} \label{sec:acknowledgements}
The work of LB and SC was supported in
part by the U.S. Department of Energy under Grant Number DE-SC0011640.
BO is supported by the National Science Foundation under Cooperative Agreement PHY-2019786 (The NSF AI Institute for Artificial Intelligence and Fundamental Interactions, \href{http://iaifi.org/}{http://iaifi.org/}).
This work also benefited from access to the University of Oregon high performance computing cluster, Talapas.

\appendix
\section{Simulation Details} \label{app:sim_details}
In this appendix, we provide further details of the simulated public datasets we use in this work \cite{Cheng:2020dal, leissner_martin_julien_2020_4641460, cheng_taoli_2021_4614656}. 
All of the QCD dijet, $W$, top, and Higgs samples are subject to the same selection criteria, showering, and detection simulation parameters. The background and anomalous events are generated using \textsc{MadGraph}~\cite{Alwall:2014hca} and \textsc{Pythia8}~\cite{Sjostrand:2014zea}, with detector effects being simulated by \textsc{Delphes}~\cite{deFavereau:2013fsa}. The jets are then clustered with \textsc{FastJet}~\cite{Cacciari:2011ma, Cacciari:2005hq} using the anti-$k_{T}$ algorithm~\cite{Cacciari:2008gp} with a cone size of $R=1.0$. All events are required to have two hard jets, with the leading jet having $p_{T}>450$ GeV and the sub-leading jet having $p_{T}>200$ GeV. We then take only the leading jet in each event.

The QCD jets are created via $pp\to jj$. The $W$ jets are created using $pp\to W^{\prime}\to W\left(\to jj\right)Z\left(\to\nu\bar{\nu}\right)$ with $m_{W^\prime}=1.2$ TeV. The top jets are produced via $pp\to Z^{\prime}\to t\bar{t}$ with $m_{Z^{\prime}} = 1.3$ TeV. Finally, the Higgs jets are produced with $pp\to HH, H\to hh, h\to jj$ with $m_{H}=174$ GeV. For each of these signals, we only consider jets with $p_{T}\in\left[550, 650\right]$ GeV. This same $p_{T}$ cut is applied to the background training and testing sets. 

\section{Network Training Hyperparameters} \label{app:training_details}
Here, we provide the details of the training hyperparameters of the AE, PNN, HLN, and isolation forests. For all three deep neural network architectures, we use the \textsc{ReduceLROnPlateau} and \textsc{EarlyStopping} callbacks from \textsc{Keras} to dynamically reduce the learning rate and stop training early, respectively. All three neural networks are trained with the Adam optimizer~\cite{Kingma:2014vow}.  

For the AE, our training hyperparameters are:
\begin{itemize}[noitemsep]
    \item{Train for 100 epochs with \textsc{EarlyStopping} on the \texttt{validation\textunderscore loss} with a patience of 10 epochs.}
    \item{Initial learning rate of $10^{-3}$ with \textsc{ReduceLRONPlateau} on the \texttt{validation\textunderscore loss} with a patience of 5 epochs.}
    \item{Batch size of 256.}
\end{itemize}

For the HLN and PNN, our training hyperparameters are:
\begin{itemize}[noitemsep]
    \item{Train for 200 epochs with \textsc{EarlyStopping} on the \texttt{validation\textunderscore loss} with a patience of 10 epochs.}
    \item{Initial learning rate of $10^{-3}$ with \textsc{ReduceLRONPlateau} on the \texttt{validation\textunderscore loss} with a patience of 5 epochs.}
    \item{Batch size of 256.}
\end{itemize}

With the early stopping conditions, the AE trains in ${\sim} 30$ epochs, the PNN trains in ${\sim} 50$ epochs, and the HLN trains in ${\sim} 60$ epochs.

For the isolation forests, our training hyperparameters are:
\begin{itemize}[noitemsep]
    \item {250 estimators in the ensemble.}
    \item {The \texttt{max\textunderscore features} used to train each estimator is set to the number of inputs for each event.}
    \item{\texttt{contamination} is set to \texttt{`auto'} since there is no way to determine what fraction of events can reliably be called outliers \emph{a priori}.}
    \item{\texttt{bootstrap} is set to \texttt{`False'}, so individual trees are trained on random subsets of the data without replacement.}
\end{itemize}

\bibliographystyle{utphys}
\bibliography{refs.bib}

\end{document}